\documentclass[useAMS,onecolumn,usenatbib]{mnras}
\usepackage[utf8]{inputenc}
\usepackage{caption}
\usepackage{amsmath,amssymb}
\usepackage{float}
\usepackage{rotating,graphicx}
\usepackage{aas_macros}
\usepackage{booktabs}
\usepackage{cleveref}
\usepackage{tabularx}
\usepackage{natbib}
\usepackage{xcolor,colortbl}
\usepackage{footnote}
\usepackage{threeparttable}
\usepackage{lscape}

\graphicspath{{./Figures/}}
\crefname{section}{§}{§§}
\crefname{section}{§}{§§}

\title[Peculiarity in IGRJ16320-4751]{Peculiar temporal and spectral features in highly obscured HMXB pulsar IGR J16320-4751 using \xmm\/}

\author[Varun et.~al.]{Varun,$^1$\thanks{E-mail: varunaries@aries.res.in} 
Nirmal Iyer,$^{2,3}$ Biswajit Paul$^4$
\\
$^{1}$Aryabhatta Research Institute of Observational Sciences, Nainital, India 263001\\
$^{2}$Department of Physics, KTH Royal Institute of Technology, Stockholm, Sweden 10691\\
$^{3}$The Oskar Klein Centre for Cosmoparticle Physics, AlbaNova University Centre, Stockholm, Sweden 10691\\
$^{4}$Raman Research Institute, Bangalore, India 560080\\
}

\makesavenoteenv{tabular}
\makesavenoteenv{table}
\begin{document}
\newcommand{\ka}{$K_{\alpha }$ }
\newcommand{\fkb}{$K_{\beta }$ }
\newcommand{\ly}{$Ly \alpha $ }
\newcommand{\gxo}{GX1+4}
\newcommand{\he}{Her X-1}
\newcommand{\smx}{SMC X-1}
\newcommand{\lmx}{LMC X-4}
\newcommand{\src}{IGR J16320-4751}
\newcommand{\mas}{IGR J16318-4848}
\newcommand{\vel}{Vela X-1}
\newcommand{\gxt}{GX 304-1}
\newcommand{\aof}{A 0535+26}
\newcommand{\gro}{GRO J1008-57}
\newcommand{\voz}{V 0332+53}
\newcommand{\oao}{OAO 1657-415}
\newcommand{\fuf}{4U 1538-522}
\newcommand{\xmm}{\textit{XMM-Newton}}
\newcommand{\rgs}{\textit{XMM-RGS}}
\newcommand{\hetg}{\textit{Chandra-HETG}}
\newcommand{\rxte}{\textit{RXTE}}
\newcommand{\cndra}{\textit{Chandra}}
\newcommand{\integral}{\textit{INTEGRAL}}
\newcommand{\bat}{\textit{Swift-BAT}}
\newcommand{\swi}{\textit{Swift}}
\newcommand{\rasm}{\textit{RXTE-ASM}}
\newcommand{\rcol}{\textcolor{red}}
\newcommand\nicom[1]{(\textcolor{violet}{\textit{NI: #1}})}
\newcolumntype{t}{>{\columncolor{red}}c}

\maketitle

\begin{abstract}
    \src\ is a highly obscured HMXB source containing a very slow neutron star ($P_{spin}\sim1300$ sec) orbiting its supergiant companion star with a period of $\sim$9 days. It shows high column density ($N_{H}\sim2-5\times10^{23}$ $cm^{-2}$) in the spectrum, and a large variation in flux along the orbit despite not being an eclipsing source. We report on some peculiar timing and spectral features from archival \xmm\ observation of this source including 8 observations taken during a single orbit. The pulsar shows large timing variability in terms of average count rate from different observations, flaring activity, sudden changes in count rate, cessation of pulsation, and variable pulse profile even from observations taken a few days apart. We note that \src\ is among a small number of sources for which this temporary cessation of pulsation in the light curve has been observed. A time-resolved spectral analysis around the segment of missing pulse shows that variable absorption is deriving such a behavior in this source. Energy resolved pulse profiles in 6.2-6.6 keV band which has a partial contribution from Fe K$_\alpha$ photons, show strong pulsation. However, a more systematic analysis reveals a flat pulse profile from the contribution of Fe K$_\alpha$ photons in this band implying a symmetric distribution for the material responsible for this emission. Soft excess emission below 3 keV is seen in 6 out of 11 spectra of \xmm\ observations.
\end{abstract}


\section{Introduction}

The study of high mass X-ray binaries (HMXB) gives us insight into many areas of modern astrophysics including accretion physics, stellar wind, stellar evolution, and formation of double compact binaries. HMXBs were among the first astrophysical sources studied using space based X-ray telescopes. All sky surveys conducted before 2000 discovered 110 HMXBs \citep{2000A&AS..147...25L} using soft X-ray instruments. A majority of these sources were Be binaries and only a few systems were classified as supergiant binaries. However, there was a systematic bias in this sample due to the operating energy range ($<$12 keV) of these surveys. The sample of HMXBs has been updated in the past two decades with hard X-rays ($>$17 keV) surveys conducted by \integral\ and \swi\  observatory. In particular, the number of supergiant HMXBs increased by almost a factor of three \citep{2015A&ARv..23....2W} and a large number of these were characterized as highly obscured HMXBs. This previously undetected subclass of HMXB shows large intrinsic photoabsorption  ($N_{H}>10^{23}$ $cm^{-2}$) in their spectra \citep{2006ApJ...644..432Z,2006MNRAS.366..274R}. \mas\ is the most extreme case in this subclass with the highest known intrinsic absorption ($N_{H} \sim 10^{24}$ $cm^{-2}$) \citep{2007A&A...465..501I}. Most of the highly obscured sources have short orbital periods and possibly undergoing a transition to Roche lobe overflow \citep{2015A&ARv..23....2W}. Highly obscured sources \src\ and EXO 1722-363 have larger orbital period ($\sim$10 days) and hence are unlikely to be in the Roche lobe transitioning phase. \\

\src\ was discovered in February 2003 by IBIS/ISGRI detector \citep{2001ESASP.459..591L} during a target of opportunity (ToO) observation of another binary source 4U 1630-47 \citep{2003IAUC.8076....1T}. It was revealed later that this source was associated with ASCA source AXJ1631.9-4751, observed in 1994 \citep{2001ApJS..134...77S}. Its X-ray position was determined accurately at RA = $16^{h}32^{m}01^{s}.9$, Dec = $-47^{\circ}52\arcmin27\arcsec$ using \xmm\/ observations \citep{2003A&A...407L..41R}. \xmm\ observations also revealed timing variability in the source and X-ray pulsation of $P=1309\pm40$ s associated with the spin period of the neutron star (NS) in the system \citep{2005A&A...433L..41L}. This binary is located at a distance of $\sim$3.5 kpc \citep{2013A&A...560A.108C}. A strong modulation at a period of $8.96\pm0.01$ days in the long term \bat\ corresponds to the orbital period of NS around its companion star \citep{2005ATel..649....1C}. The presence of several emission lines in the near infra-red (NIR) spectrum (e.g. Pa(7-3), Br(17-4), and HeI(7-4)) indicate that the companion star might be an O/B supergiant \citep{2008A&A...484..783C}. Companion star is not visible in any optical bands corresponding to the X-ray position of the source due to the high intrinsic absorption.\\

X-ray light curves of IGR~J16320-4751 show large variability at various time scales, from seconds-minutes \citep{2003A&A...411L.373R} to days-months \citep{2004ESASP.552..247F}. The pulse profile and fraction are also seen to vary with time. However, the pulse fraction is found to be constant with energy in both \xmm\ and \integral\ energy bands \citep{2006MNRAS.366..274R}.  The X-ray spectrum of this source is similar to other highly absorbed HMXBs, consisting of a power-law, prominent iron emission lines along with a large absorption column ($N_H\simeq2-5\times10^{23} cm^{-2}$). Despite this large column density, the presence of an excess of soft photons below $\sim$3 keV has been noted \citep{2006MNRAS.366..274R}. The absorption, $N_{H}$, and line emission (EW of Fe $K_{\alpha}$) are correlated with each other, as expected from a system with the absorbing material also yielding fluorescence lines \citep{2018A&A...618A..61G}. \\

Attempts have been made before to understand the orbital variations of IGR~J16320-4751 by proposing for and studying multiple observations along a single orbit~\citep{2009AIPC.1126..313H,2018A&A...618A..61G}. In this work, we use these observations to study the variable pulse profile of the neutron star. Iron emission lines have been found in spectra of a large number of HMXBs and have proven to be a useful tool to study stellar wind properties in these systems~\citep{2010ApJ...715..947T,2018A&A...610A..50P,2017xru..conf...28A}. We study the pulse profile in the energy range of these iron line emissions to look for clues about the overall pulsation behavior. From these studies, we gather evidence for some peculiar temporal and spectral features in this binary system. These studies are detailed in the following sections. We give details of the instruments and observations used in \cref{sec:obsec}. Details on the timing and spectral analysis undertaken are given in \cref{sec:ana}. Finally, we present our results followed by a discussion of these findings in \cref{sec:dis}. \\

\section{Observation and data reduction}
\label{sec:obsec}
We used archival data from \xmm\ observations of \src\/ for this work. \xmm\ is a space-based high energy mission launched by the European Space Agency (ESA)). A set of three European Photon Imaging Cameras (EPIC) serve as the primary instruments on board \xmm\ \citep{2001A&A...365L...1J}. The EPIC cameras consist of a single PN CCD camera and two MOS (Metal Oxide Semi-Conductor) CCD cameras. The PN camera is made of twelve CCD units embedded into a single silicon chip. The EPIC PN camera has a moderate effective area ($\sim$1000 $cm^{2}$) and good spectral resolution ($\Delta E/ E\sim20-50$) in its nominal energy range (0.15-15 keV). The observations used for this paper are done in the full-frame imaging mode, in which EPIC PN has a time resolution of 73.4 ms. IGR J16320-4751 was observed 11 times with \xmm\ between March 2004 and September 2008. Details of these observations are given in Table~\ref{obs}. The observation numbers given in Table~\ref{obs} are used to identify individual observations in the remainder of this paper.  8 of 11 observations (Obs \#3 - Obs \#10) used in this work were carried out during 14-26 August 2008 at different orbital phases of the same binary orbit. This allows us to investigate orbital variations in detail over the same orbit.\\

\begin{table}
    \centering
    \caption{Table Of Observations}
    \label{obs}
    \begin{tabular}{||c|c|c|c|c|c|c|c||}
     \hline
     Obs.No. & Obs. Id 	& Date-Start & 	Tstart	 & 	Date-End   & 	Tend 	 & Duration[s]  &  Orbit Phase \\
     \hline
      1 & 0128531101    & 04-03-2004 &  20:58:26 &  05-03-2004 &    03:12:27 &  22441       & 0.769-0.797 \\
      2 & 0201700301    & 19-08-2004 &  13:28:23 &  20-08-2004 &    03:20:40 &  49337       & 0.419-0.483 \\                            
      3 & 0556140101 	& 14-08-2008 & 	22:44:10 &	15-08-2008 & 	01:14:43 & 	9033        & 0.401-0.413 \\
      4 & 0556140201 	& 16-08-2008 & 	17:41:23 & 	16-08-2008 & 	19:55:16 & 	8033        & 0.601-0.611 \\	
      5 & 0556140301 	& 18-08-2008 & 	13:36:12 &	18-08-2008 & 	15:33:26 & 	7033        & 0.804-0.814 \\
      6 & 0556140401 	& 20-08-2008 & 	07:36:07 & 	20-08-2008 & 	10:43:20 & 	11233       & 0.999-0.013 \\
      7 & 0556140501 	& 21-08-2008 & 	07:04:12 & 	21-08-2008 & 	07:41:30 & 	2238        & 0.108-0.110 \\
      8 & 0556140601 	& 22-08-2008 & 	03:55:58 & 	22-08-2008 & 	07:22:36 & 	12397       & 0.204-0.220 \\
      9 & 0556140701 	& 24-08-2008 & 	18:30:00 & 	24-08-2008 & 	21:00:33 & 	9033	    & 0.494-0.506 \\
      10 & 0556140801 	& 26-08-2008 & 	13:34:33 & 	26-08-2008 & 	16:15:06 & 	9632	    & 0.694-0.706 \\
      11 & 0556141001   & 17-09-2008 &  01:25:41 &  17-09-2008 &    03:31:14 &  7555        & 0.084-0.094 \\
     \hline
    \end{tabular}
    \begin{tablenotes}
    \item[1] {\bf Notes:} Orbital phases of different observations has been calculated using orbital period = $8.991\pm0.001$ days (obtained in this work) and taking reference epoch at the mid point of observation \#9 (54702.82310 MJD).
    \end{tablenotes}
\end{table}

To analyze the timing and spectral properties of \src\ from the EPIC PN event files, we extracted source information using a circular region of radius $40\arcsec$ centered on the source position. The analysis was done using XMM Science Analysis System (SAS version 20.0.0). Background events were extracted from a similar source-free circular region. Both light curves and spectra were extracted from these regions and the background was subtracted from the source counts. The light curves were also barycenter corrected using the \textbf{SAS} tool \textbf{barycen}. The spectra were grouped to have a minimum of 25 counts in each bin with \textbf{specgroup} and fit using XSPEC V12.10.1.\\

\begin{figure}
    \centering
    \includegraphics[trim=1cm 0.8cm 3cm 0.5cm, clip=true, width=0.54\textwidth]{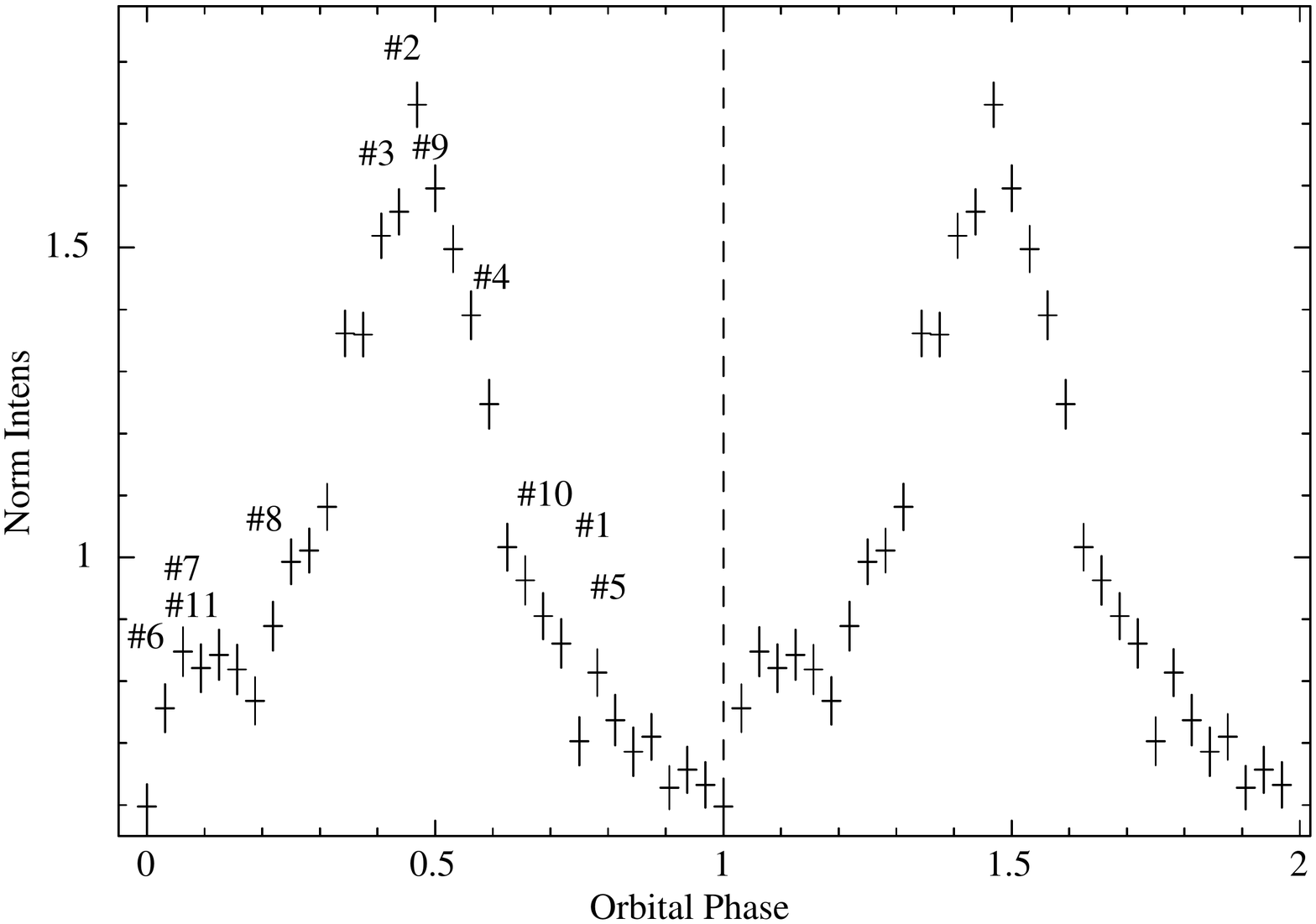}
    \includegraphics[trim=4cm 0.5cm 5cm 0.5cm, clip=true, width=0.43\textwidth]{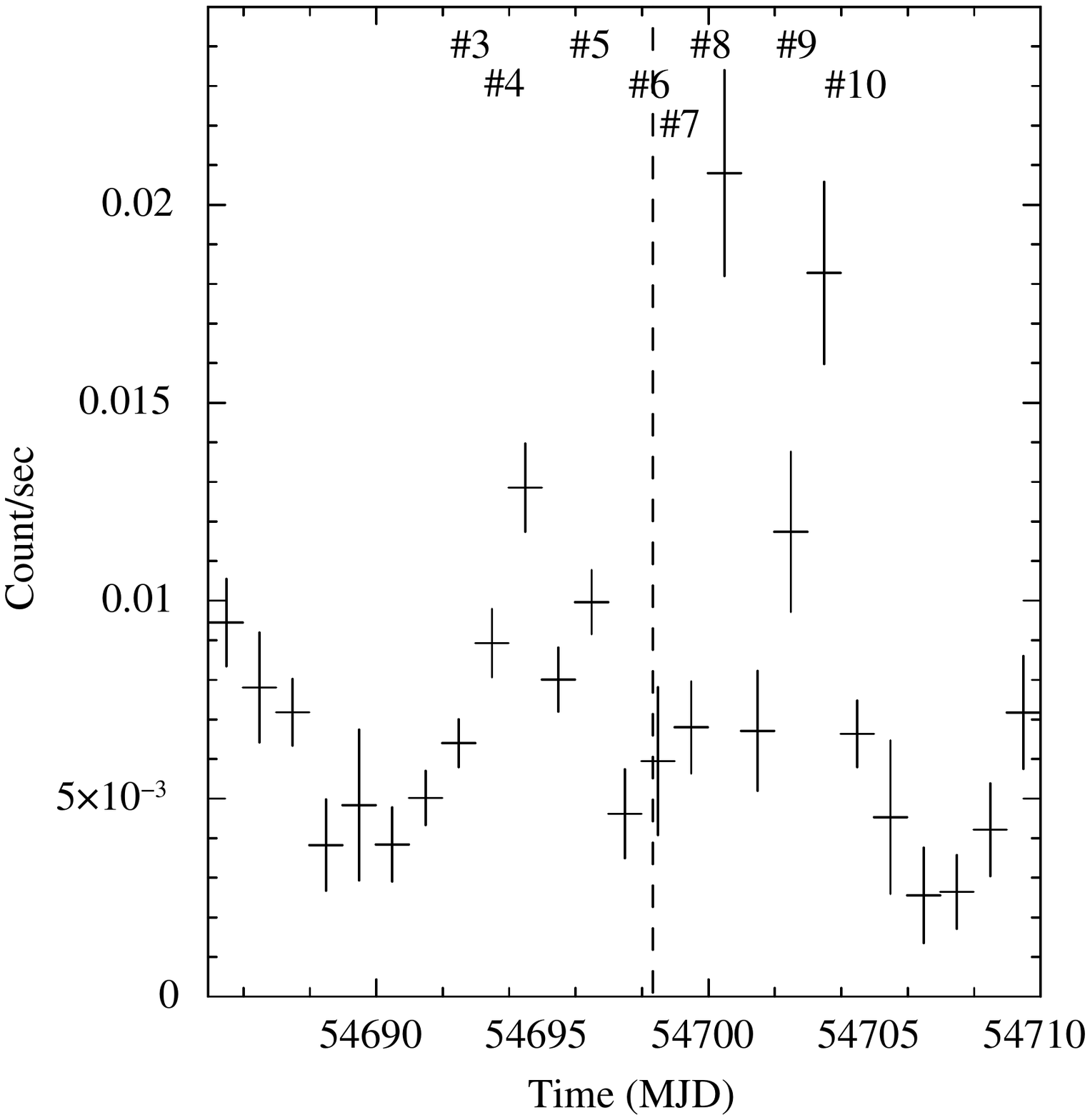}
    \caption{\emph{Left panel:} Orbital profile of \src\/ using long term \bat\ data. Two cycles of orbital profile are shown on which orbital phases of all \xmm\ observations are marked in the first cycle. \emph{Right panel:} Actual \bat\ light curve for August 2008 during which 8 \xmm\ observations were taken at different phases covering the entire orbit. The vertical dotted line in this panel corresponds to orbital phase 0.}
    \label{orpr}
\end{figure}

We used the \bat\/ long term light curve from February 2005 to December 2020 and searched for periodicity to determine a period of $8.991\pm0.001$ days. The error on the spin period value is found bootstrap sample of 1000 light curves \citep{2004AstL...30..824F,2013AstL...39..375B}. We generated copies of original light curves using a Gaussian random number generator taking each data point with value and error on it as mean and standard deviation respectively. The best period was determined from each light curve using the tool \textbf{efsearch}. Finally, we computed the standard deviation of the distribution of period values. This value is similar to the orbital period reported in \citep{2018A&A...618A..61G}, with a lower error owing to a longer dataspan. The orbital profile created with this period folded using a reference epoch at 54702.82310 MJD is shown in the left panel of Figure \ref{orpr}. It is interesting to note that although the source is not an eclipsing HMXB system, it shows a very strong orbital intensity modulation. The \xmm\ observations we use in this paper, are marked on the orbital profile in Fig.~\ref{orpr} at their respective orbital phase values. The actual \bat\/ light curve during the single binary orbit covering 8 \xmm\ observations in August 2008 is shown on the right panel of Figure \ref{orpr}.

\section{Analysis and Results}\label{sec:ana}

\subsection{Timing}
\begin{figure}
    \centering
    \includegraphics[scale=0.5]{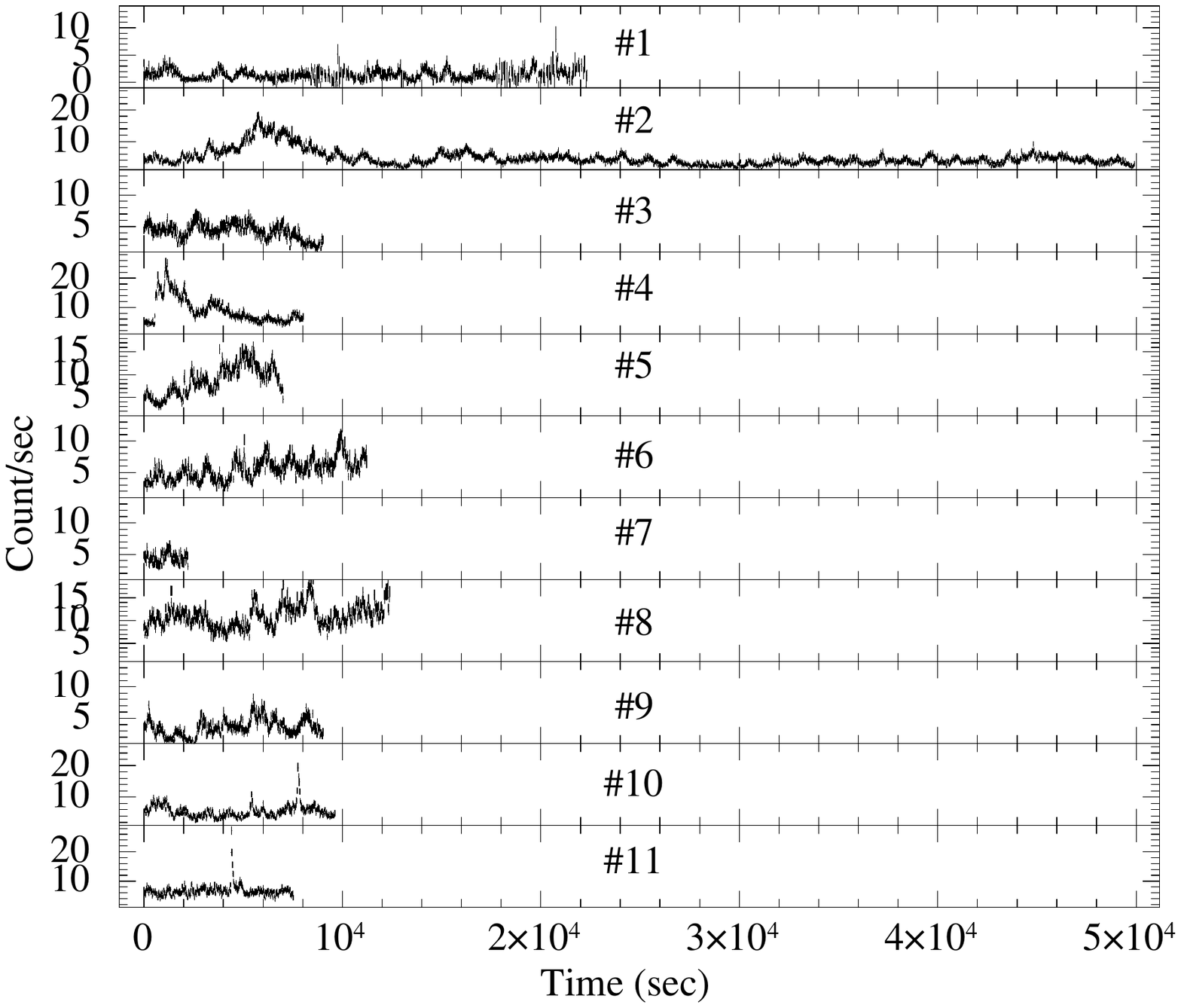}
    \caption{Light curves from all \xmm\ observations in Table \ref{obs} are plotted with 20.0 sec time bins. Time axis has been rescaled to start from zero for all observations.}
    \label{lcs}
\end{figure}

Pulsations are immediately apparent in the EPIC PN light curves of \src. Before analyzing these pulsations, we inspected the light curves to see if any other significant features were visible. A noticeable feature is a difference in the count rate seen in the light curves from different observations. Except for observation \#1 which shows a steady non-variant light curve, all other observations display variability of some kind. Flaring activity for a few ks can be seen in observations \#2 and \#4 during which the count rate increases by a factor of $\sim$5 compared to the average count rate in the light curve (see Figure \ref{lcs}). Observations \#10 and \#11 show comparatively shorter flares (2 and 1 respectively) for a few 100 seconds. The count rate in observation \#3 is steady for 8 ks and it starts to fall in the end. The count rate in observations \#5 and \#6 show gradual increase whereas observations \#8 and \#9 shows random rise and fall.  \\  

\begin{figure}
    \centering
    \includegraphics[scale=0.18]{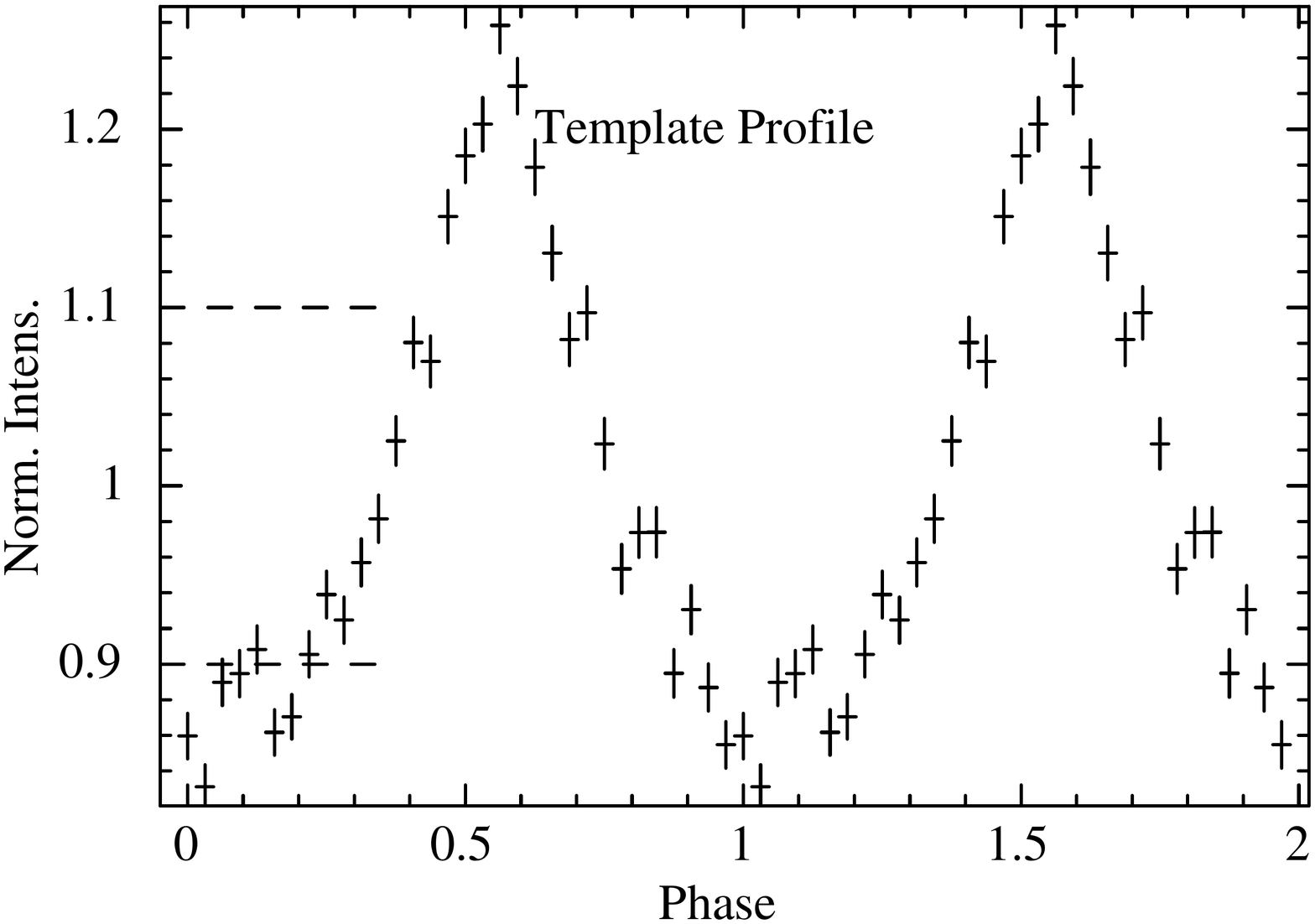}
    \includegraphics[scale=0.18]{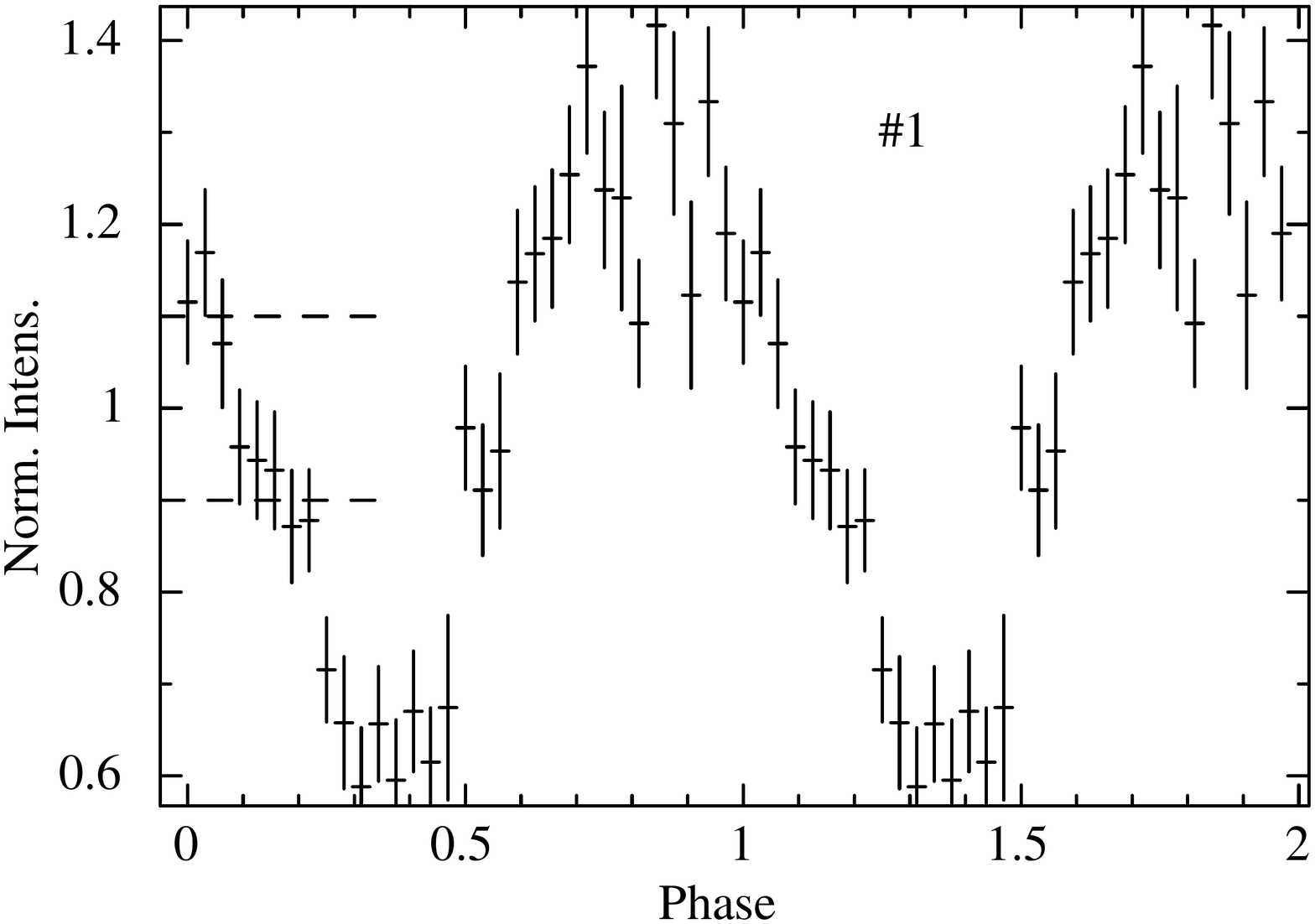}
    \includegraphics[scale=0.18]{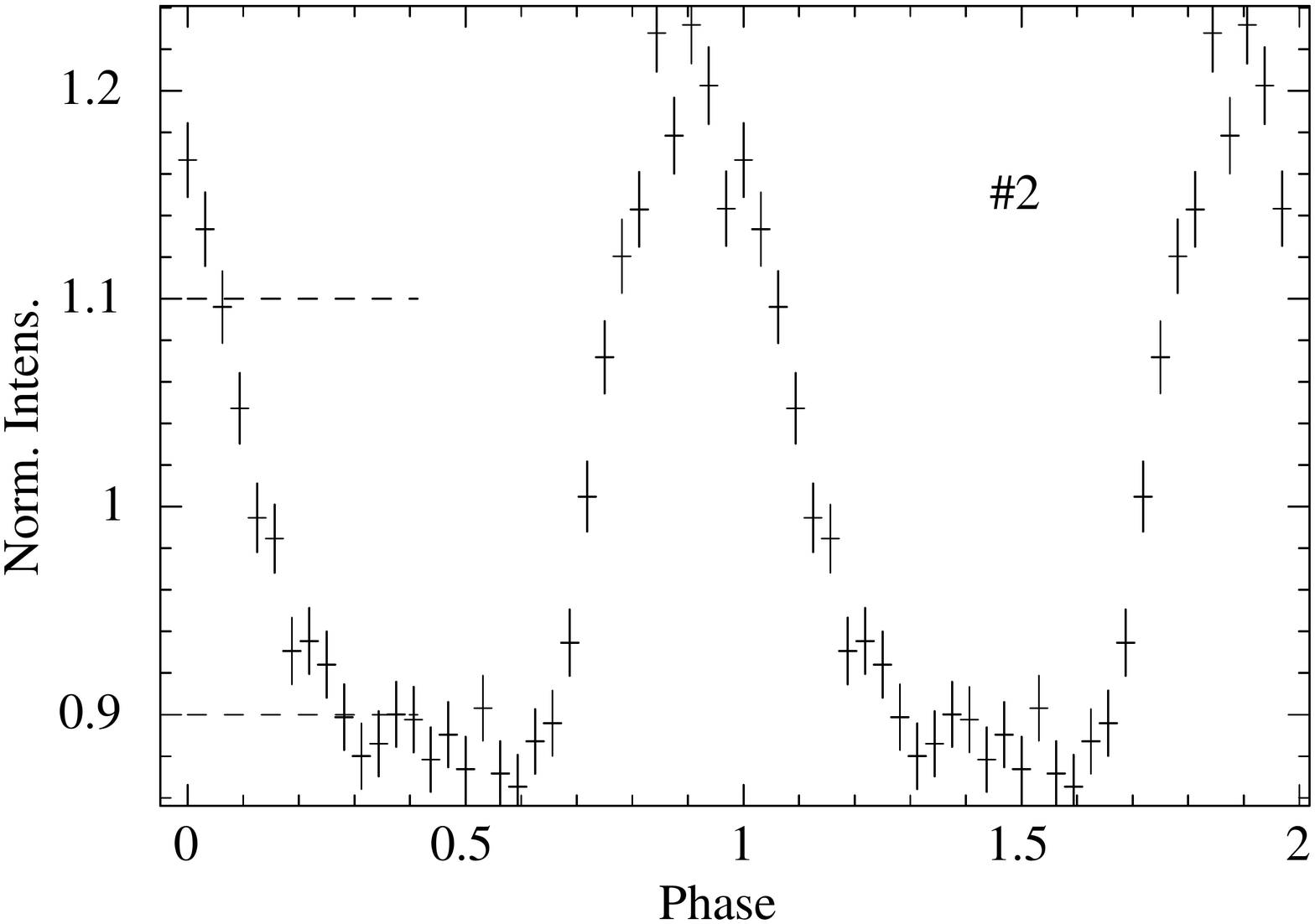}\\
    \includegraphics[scale=0.18]{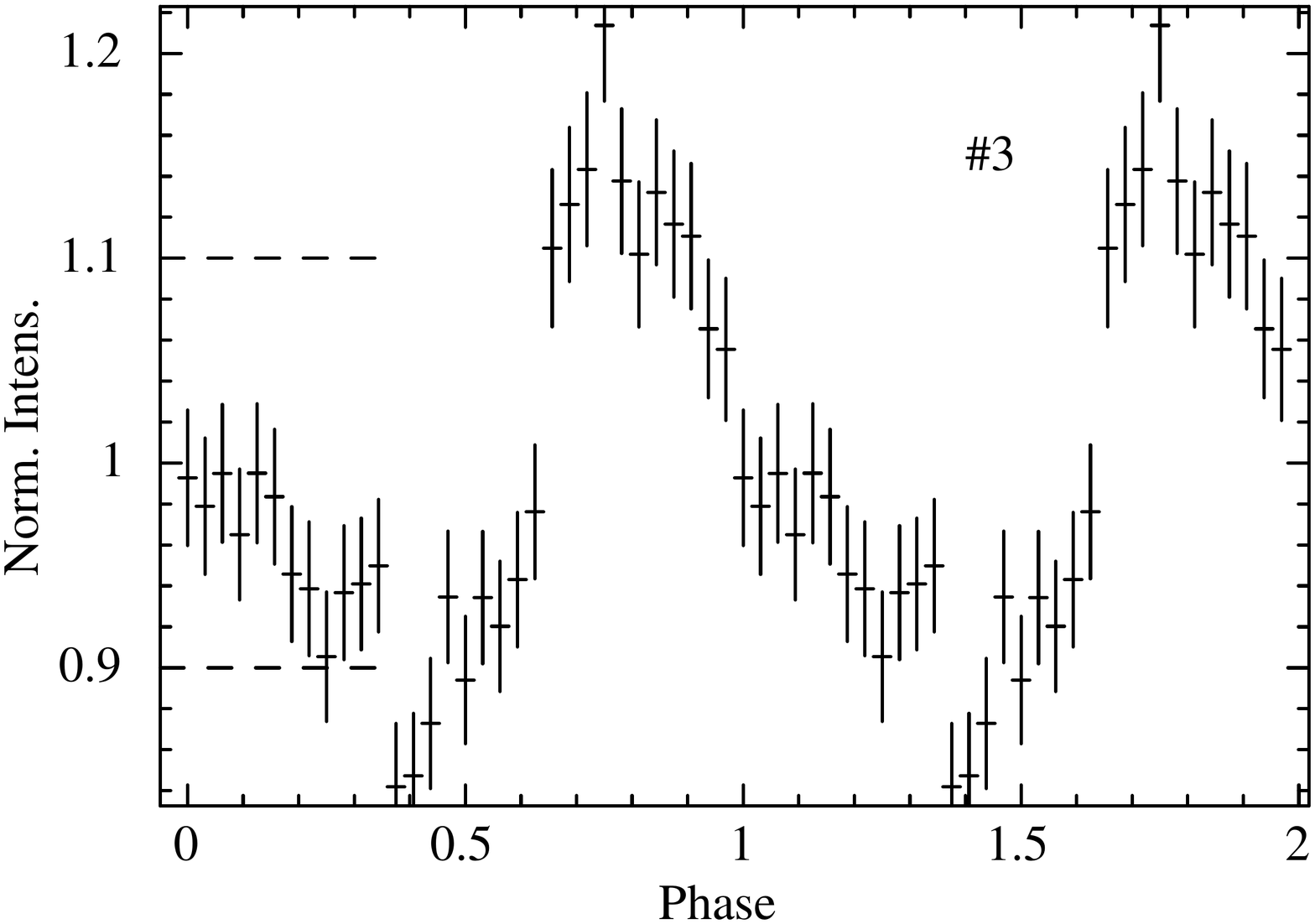}
    \includegraphics[scale=0.18]{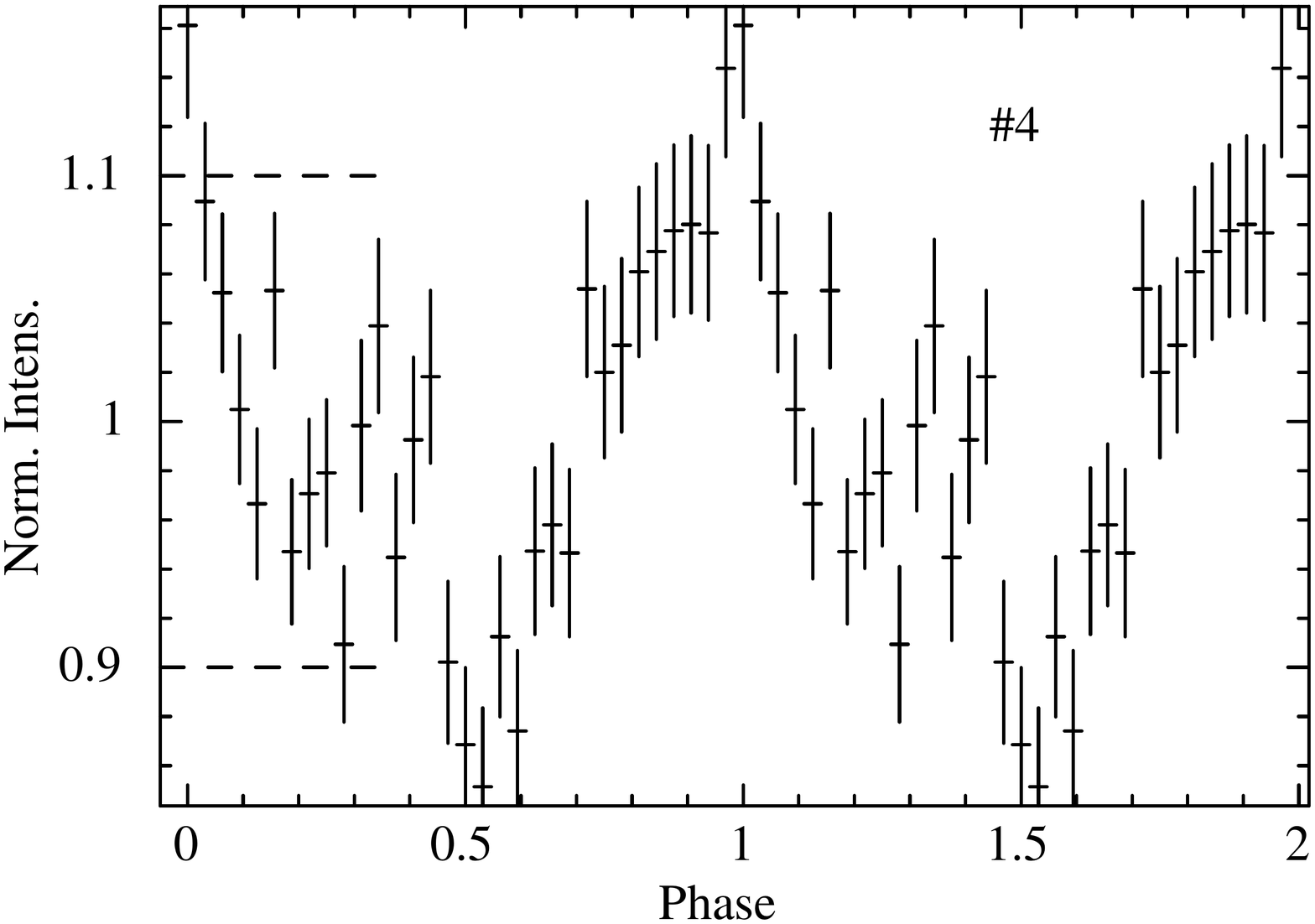}
    \includegraphics[scale=0.18]{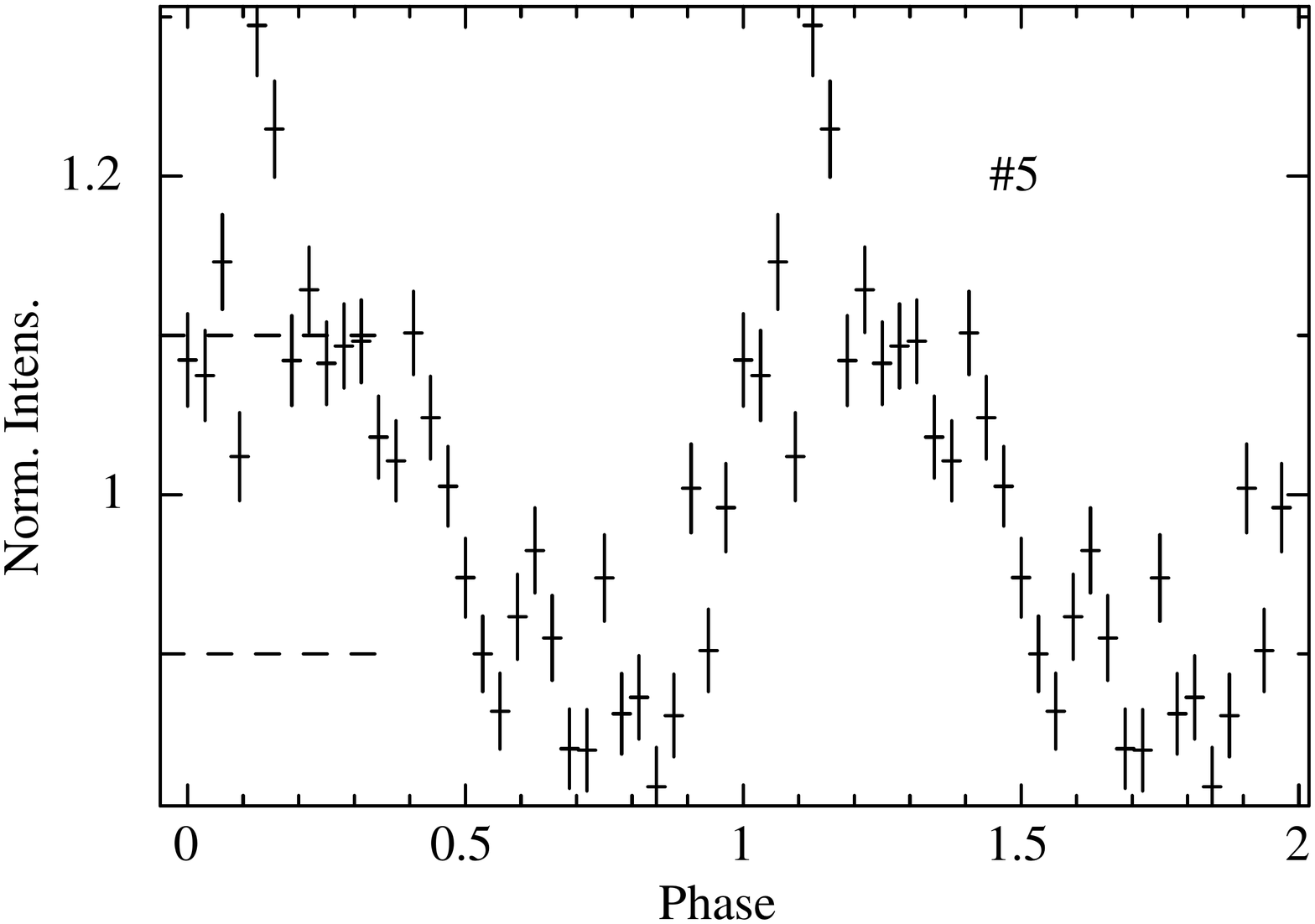}\\
    \includegraphics[scale=0.18]{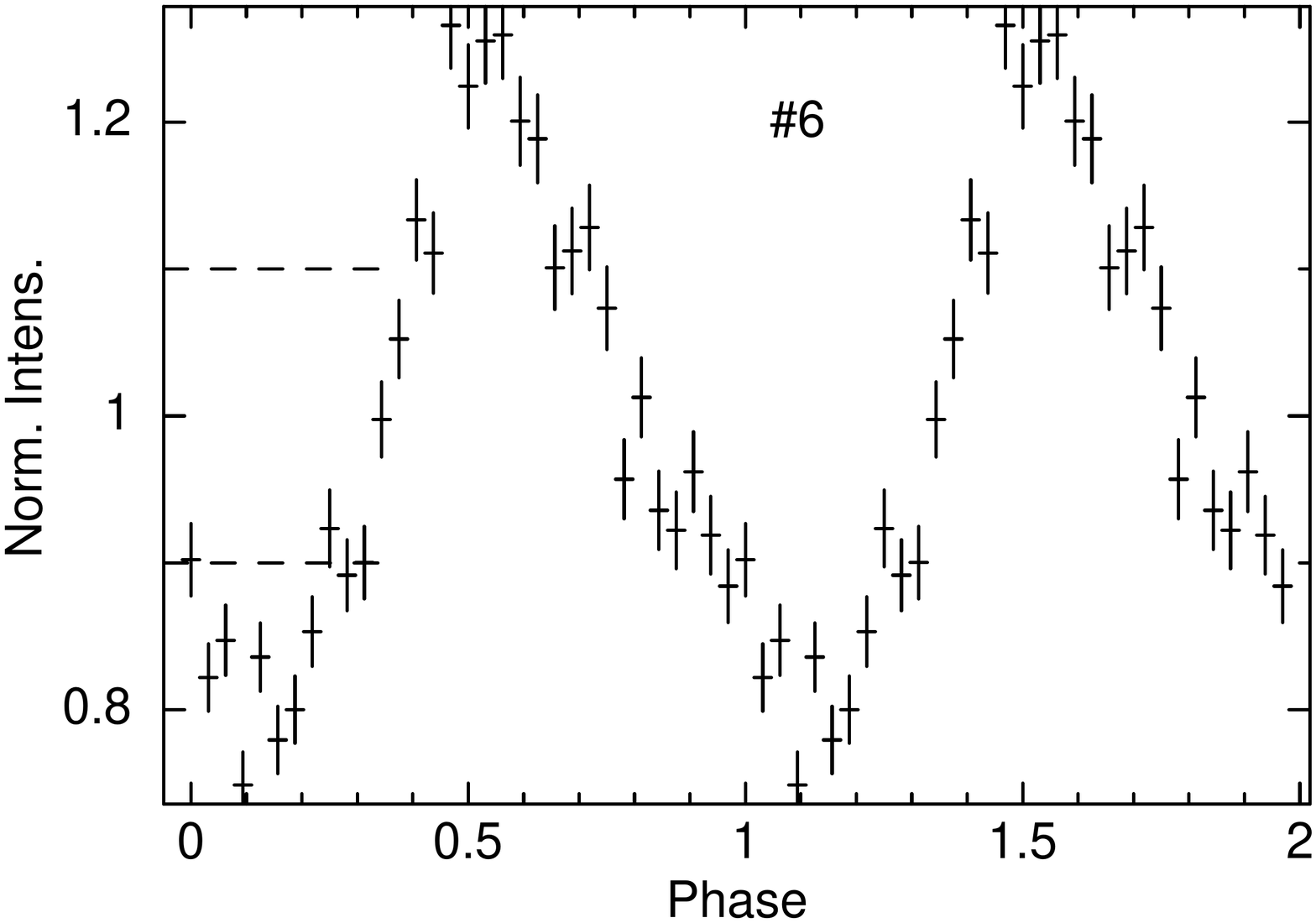}
    \includegraphics[scale=0.18]{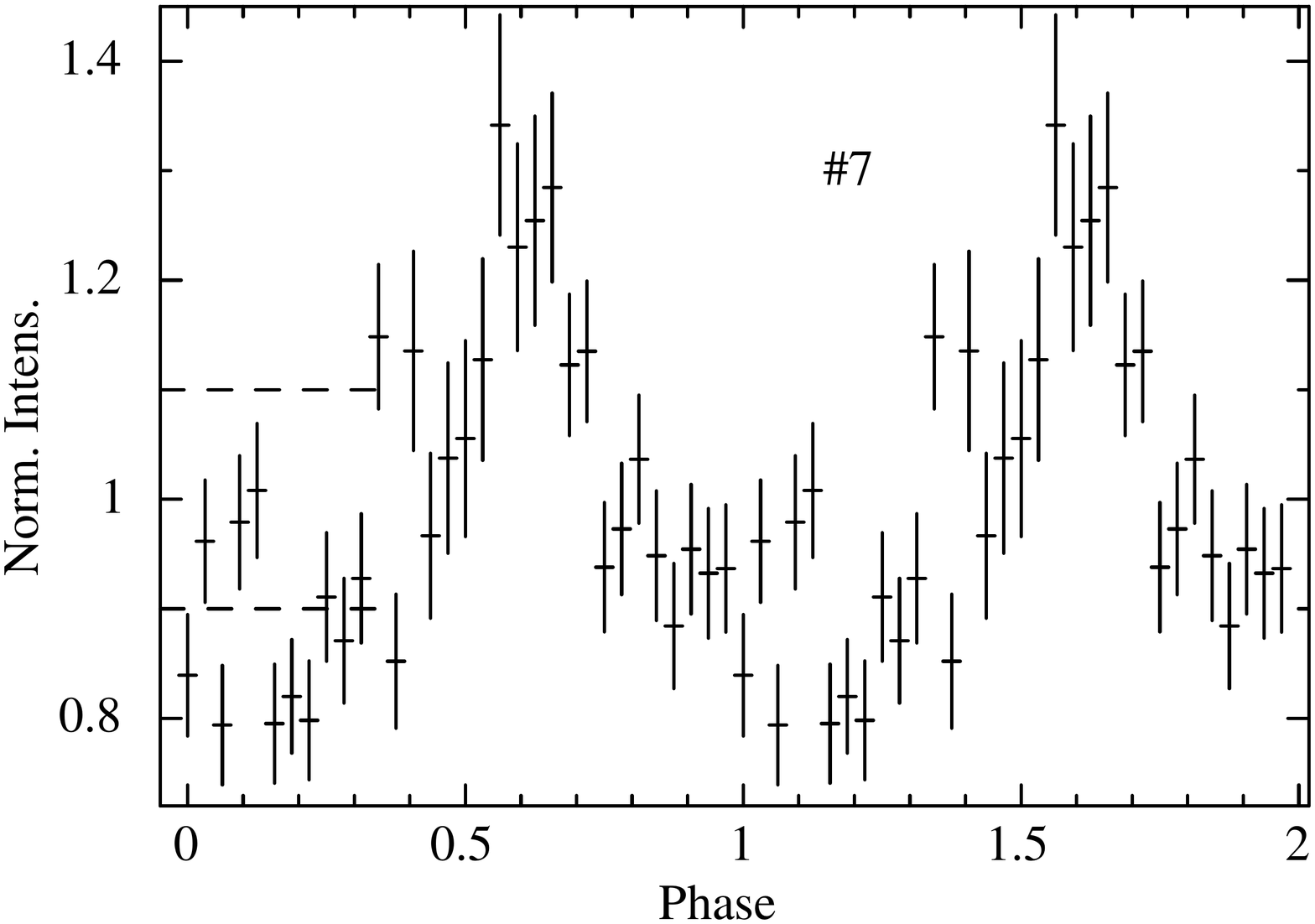}
    \includegraphics[scale=0.18]{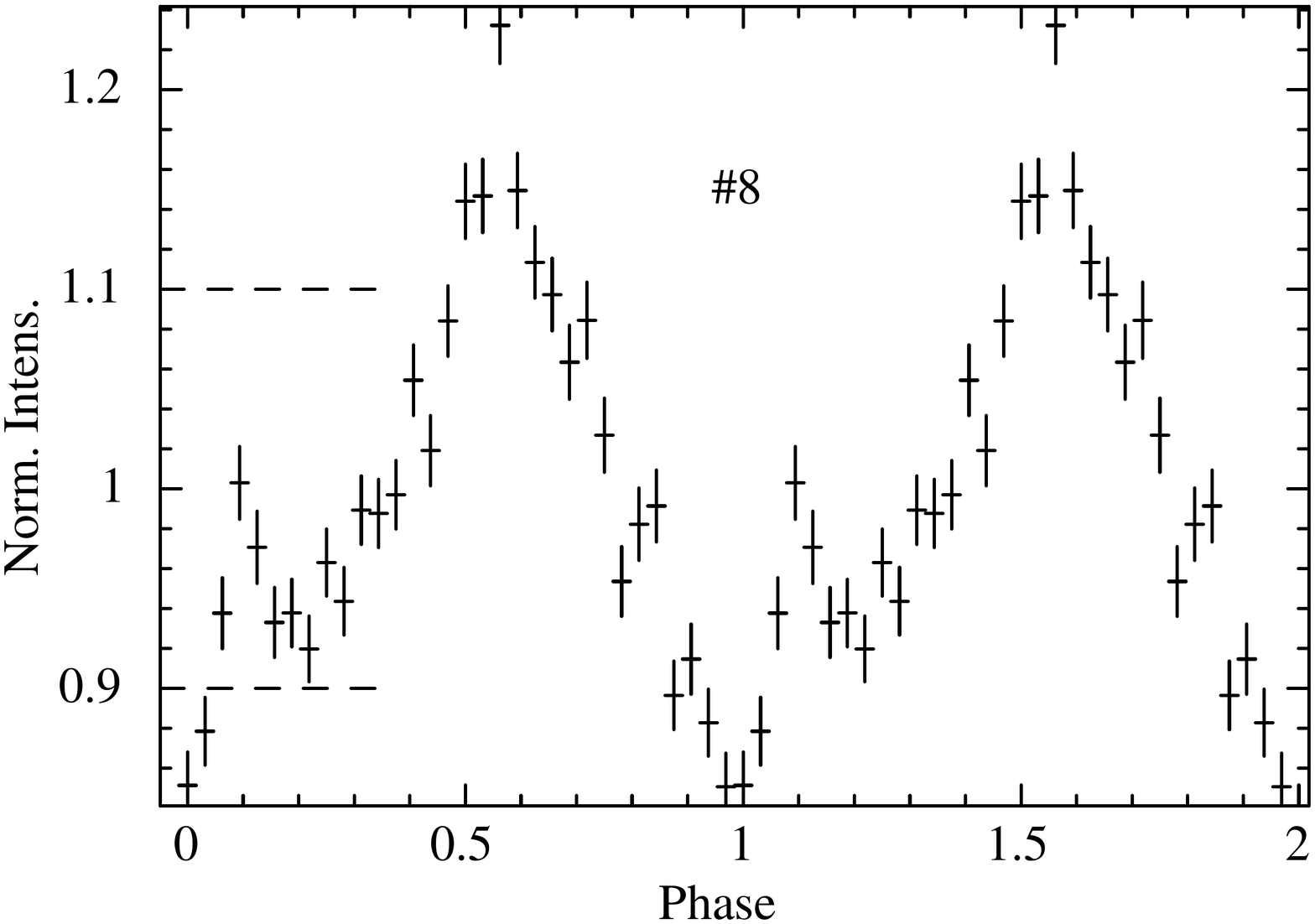}\\
    \includegraphics[scale=0.18]{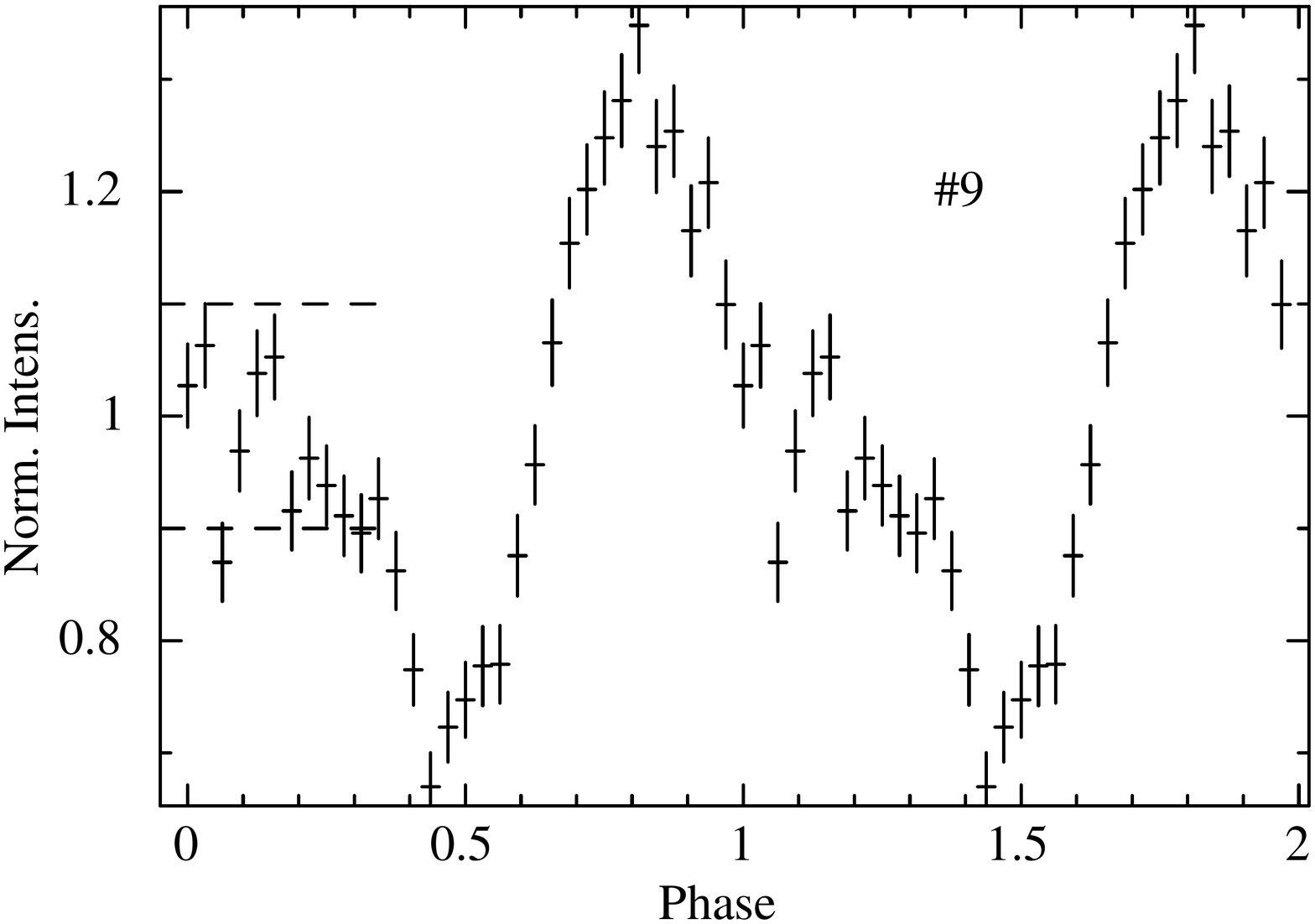}
    \includegraphics[scale=0.18]{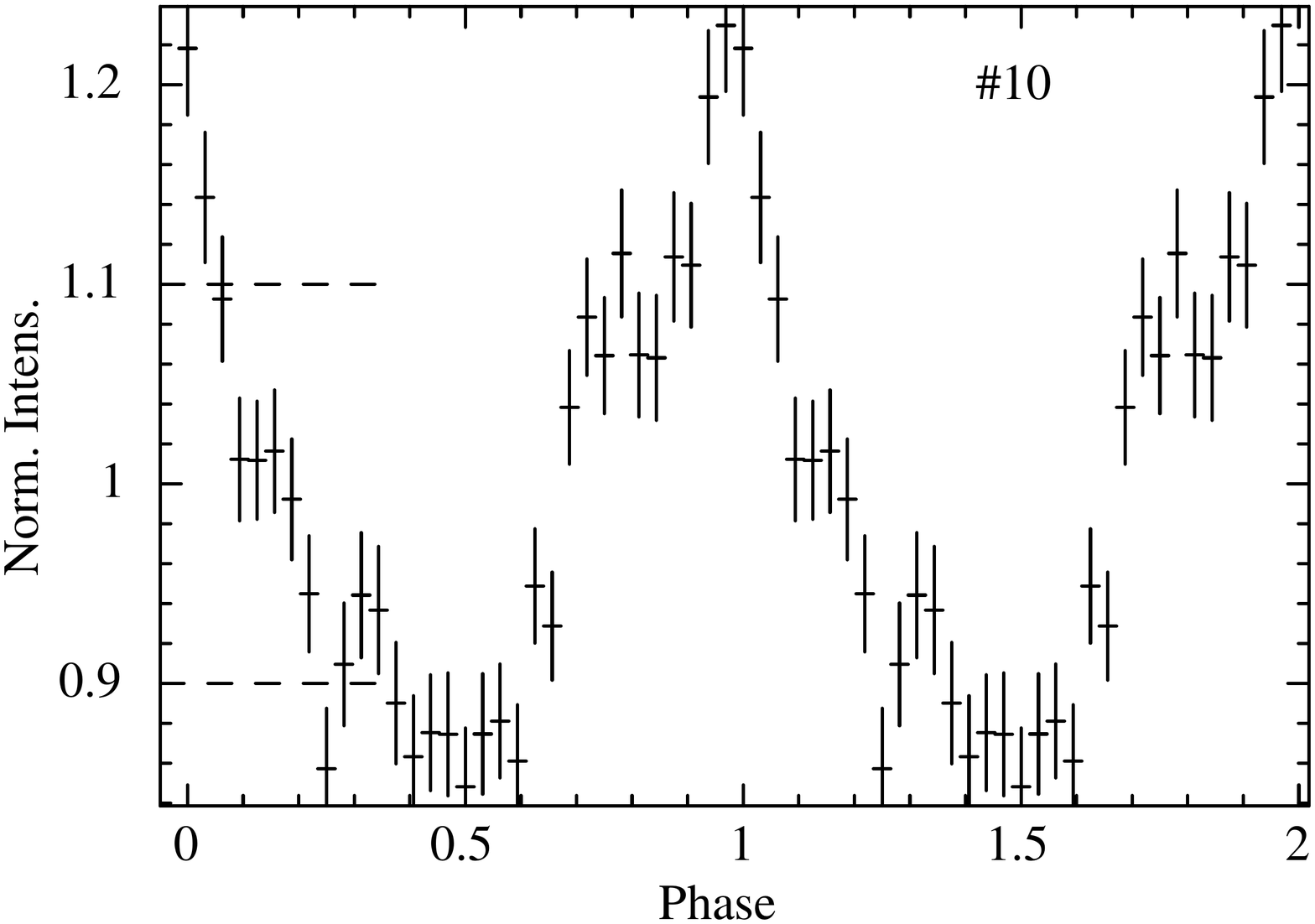}
    \includegraphics[scale=0.18]{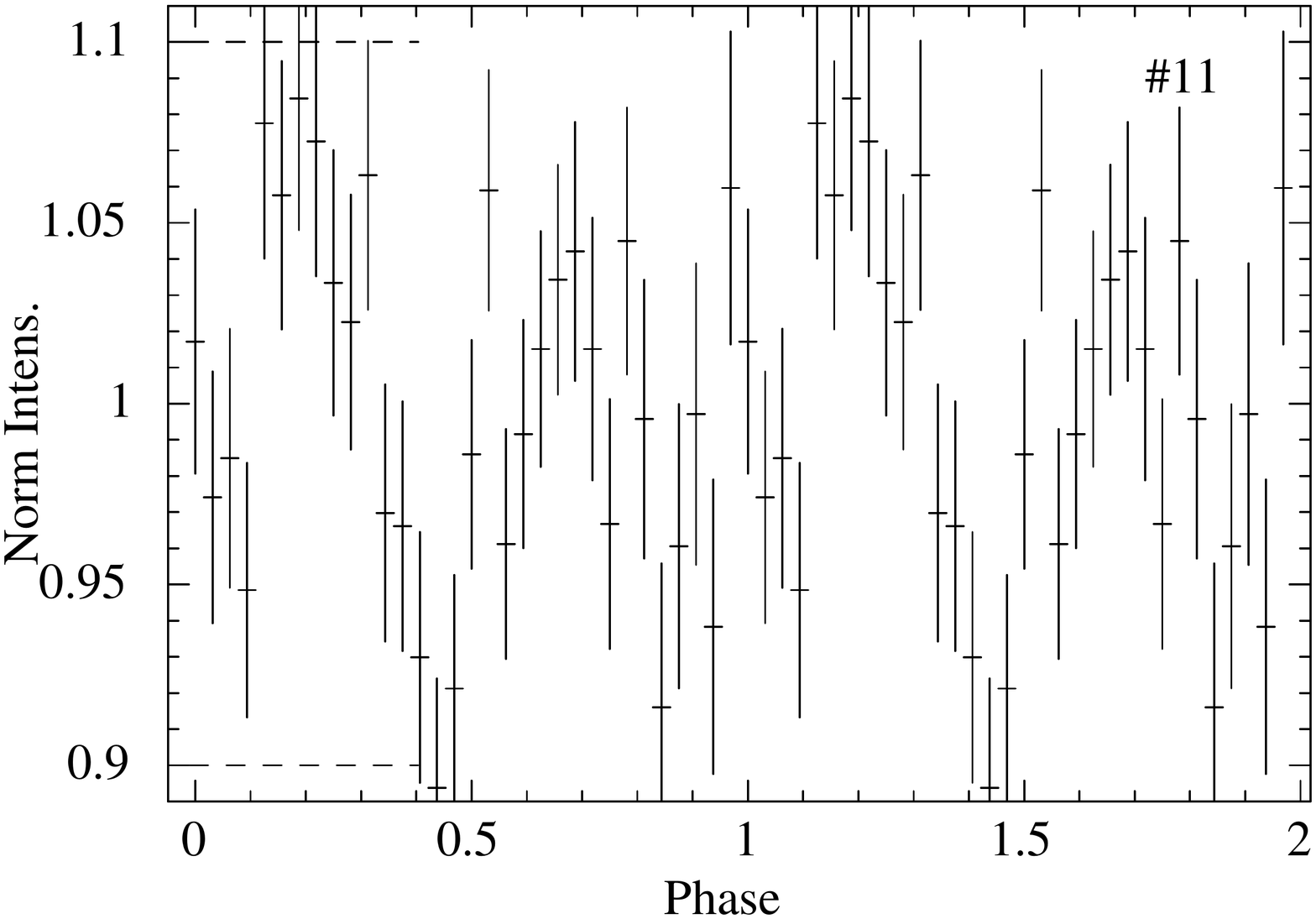}\\
    \caption{Pulse profiles obtained from different \xmm\ observations. Top left panel shows a pulse profile created from three close observations taken. Later pulse profiles are for different individual observations. Each pulse profile contains a marker for the observation number as presented in the Table \ref{obs}. Two dashed horizontal lines are marked in each pulse profile at 0.9 and 1.1 to demonstrate the variability among them.}
    \label{profiles}
\end{figure}

We observed that flares affect the shape of the pulse profile and the results of our pulse profile analysis. So we removed these sections from light curves before creating pulse profiles. We searched for periodicity in each \xmm\ observation using the tool \textbf{efsearch}. This gave us a most probable period value of 1305.9 seconds, which was obtained by using data combined from three observations made on two consecutive days, viz. obs \#6, \#7, and \#8. We created a template pulse profile by combining light curves from three observations and pulse profiles from each observation in energy band 0.2-10.0 keV using the most probable period and a reference epoch at MJD 54692.0, with the tool \textbf{efold}. All pulse profiles (created with 32 phasebins) are shown in Figure~\ref{profiles}. We note that the pulse profiles show a broad single peaked feature. The pulse profile of obs~\#8 also shows a small inter-pulse. Given that 8 of 11 observations were made during a single orbit, we attempted to use the pulse profiles to determine the orbital characteristics of the pulsar using phase lags between observations. The template profile is correlated against the profiles from individual observations to find the phase lag between observations. As seen from the pulse profiles in Figure~\ref{profiles}, there is a significant shift in the position of the primary pulse peak (for example between obs \#5 and obs \#8) which cannot be easily explained in terms of orbital changes alone. For the orbital motion to lead to such large phase-lags ($\sim$0.4), would require an extremely large mass function $\sim$280$M_\odot$. Apart from orbital motion, a change in pulsar period can also lead to shift in the phases of pulse profiles if all the light curves are folded with a constant period. The estimated period derivative to derive a phase shift of $\sim$0.4 comes out to be 5$\times10^{-4}$ ss$^{-1}$ which is much larger than the period derivatives of 10$^{-12}$ ss$^{-1}$ seen for HMXBs \citep{2014MNRAS.437.3664H}. It is more likely that these phase-lags are caused by other reasons as elaborated in the discussion section. \\

\begin{figure}
    \centering
    \includegraphics[scale=0.29]{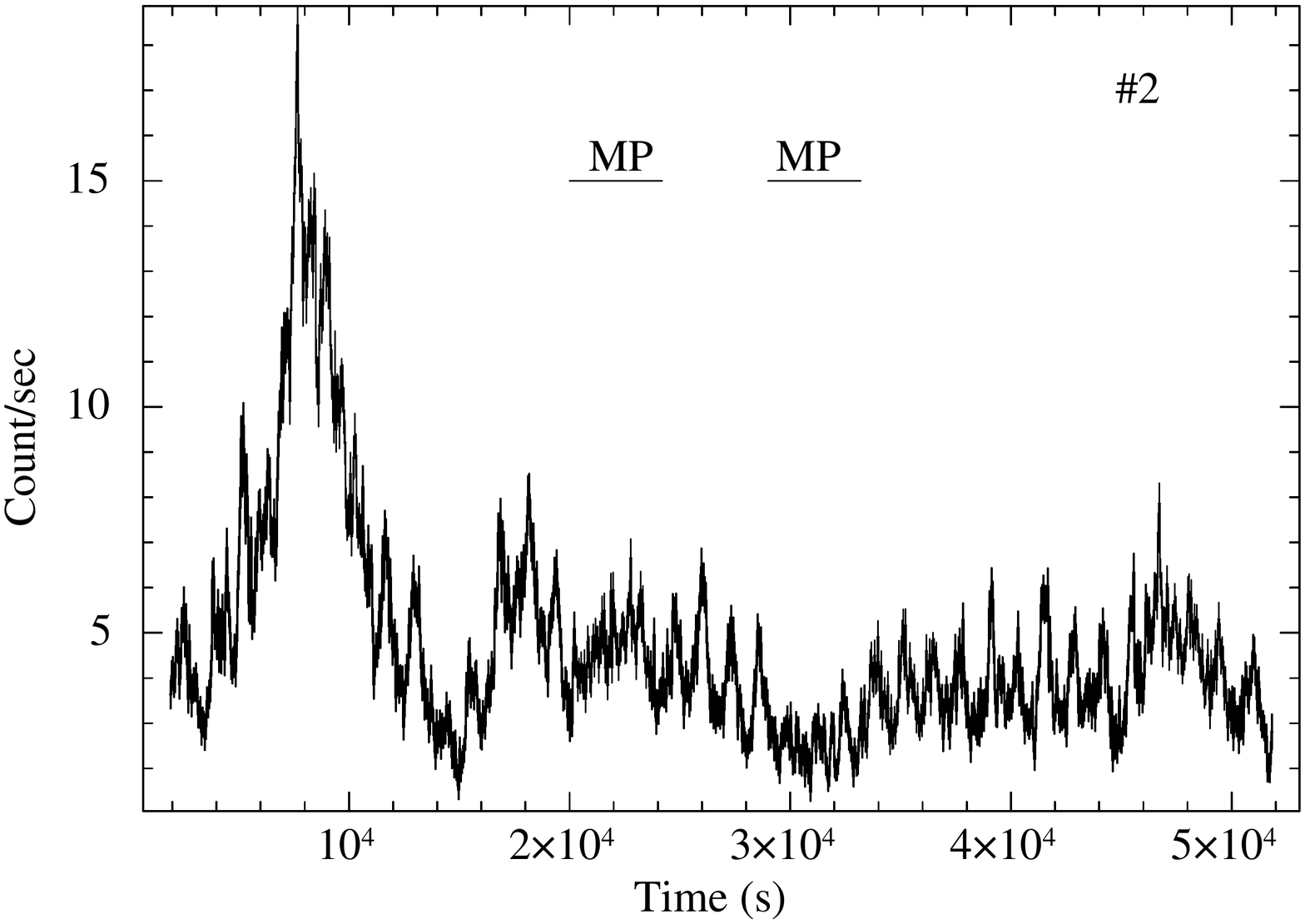}
    \includegraphics[scale=0.29]{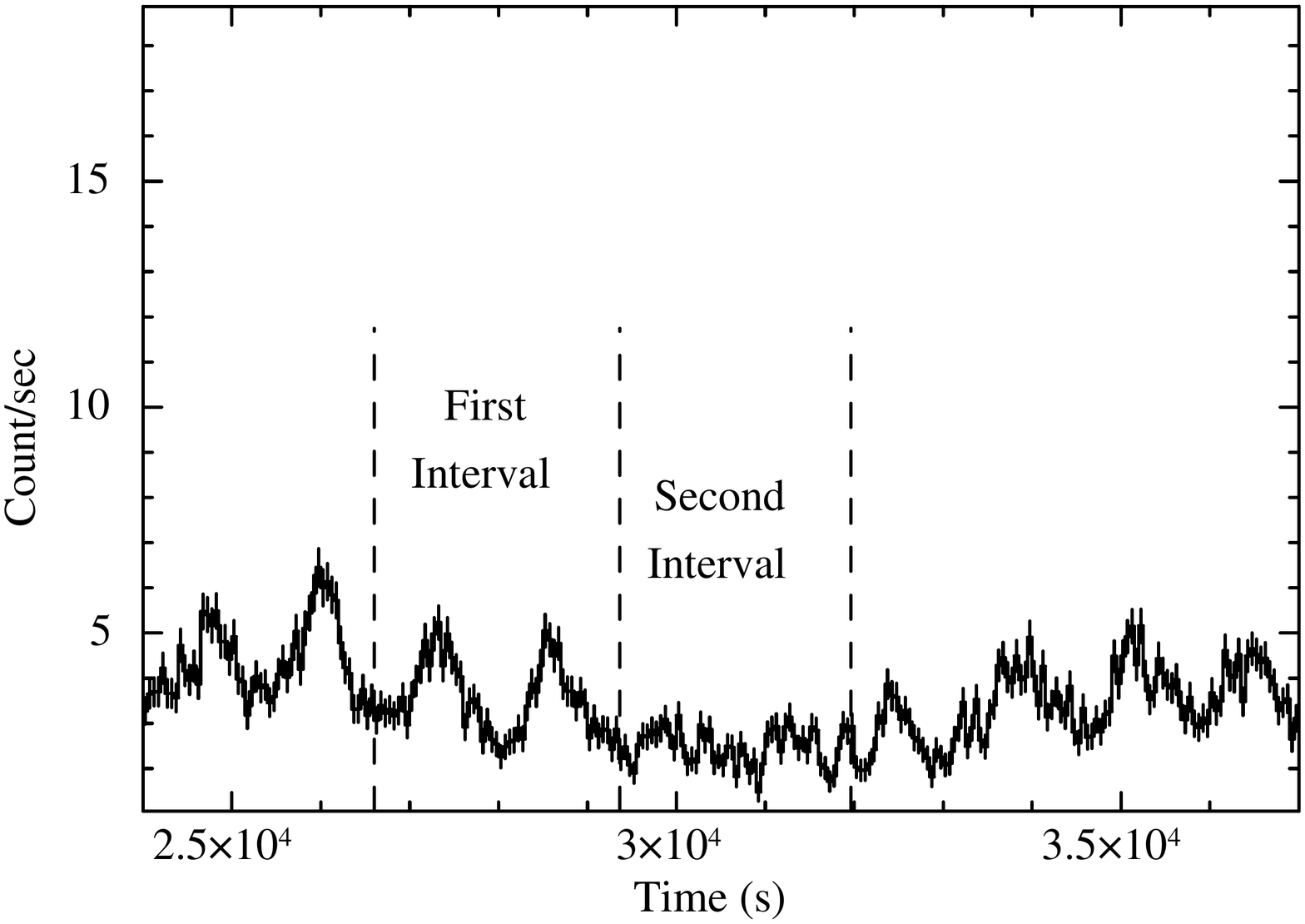}
    \caption{\emph{Left Panel:} Light curve from observation \#2 of \src\ using 20 s binning time. Time intervals during which pulses are missing are marked as MP.\emph{ Right Panel:} An expanded view near 30 ks in this observation. First interval has clear pulsation whereas two pulses are missing continuously in the second interval.  }
    \label{fig:mp}
\end{figure}

We noted a curious absence of pulsations in obs \#11. The pulsations are seen as distinct $\sim$1300 s modulations in all light curves except obs~\#11. We also noticed many instances of pulse to pulse variations in other observations where the characteristic pulse is either missing or significantly different from the broad single peak. We investigated this in more detail by overlaying the pulse profile created from the observation over the entire duration of the light curve and found pulse-to-pulse variation in 6 out of 11 \xmm\ observations of \src\/. There are several instances of single missing pulses as well as multiple missing pulses together. In Figure \ref{fig:mp}, we show the light curve of observation \#2 of \src\/. Missing pulse intervals are marked in the light curves as ``MP" (Left panel). An expanded view near 30 ks in this observation is shown to display this more clearly where the first interval has clear pulses and two pulses are missing in the second interval (Right panel). A comparison of spectral parameters between these two intervals is given in the spectroscopy section. \\

\begin{figure}
    \centering
    \includegraphics[width=0.45\textwidth]{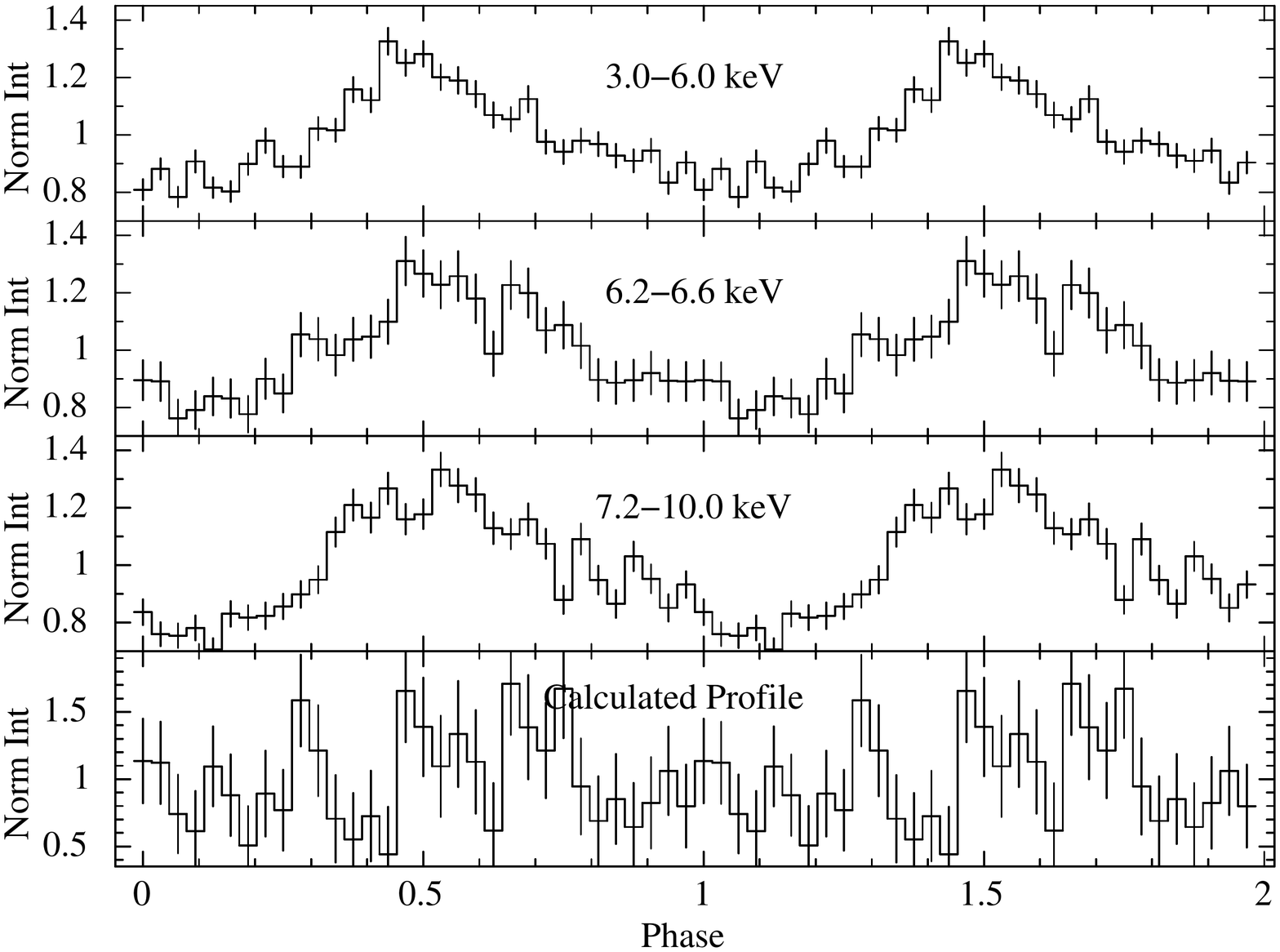}
    \includegraphics[width=0.45\textwidth]{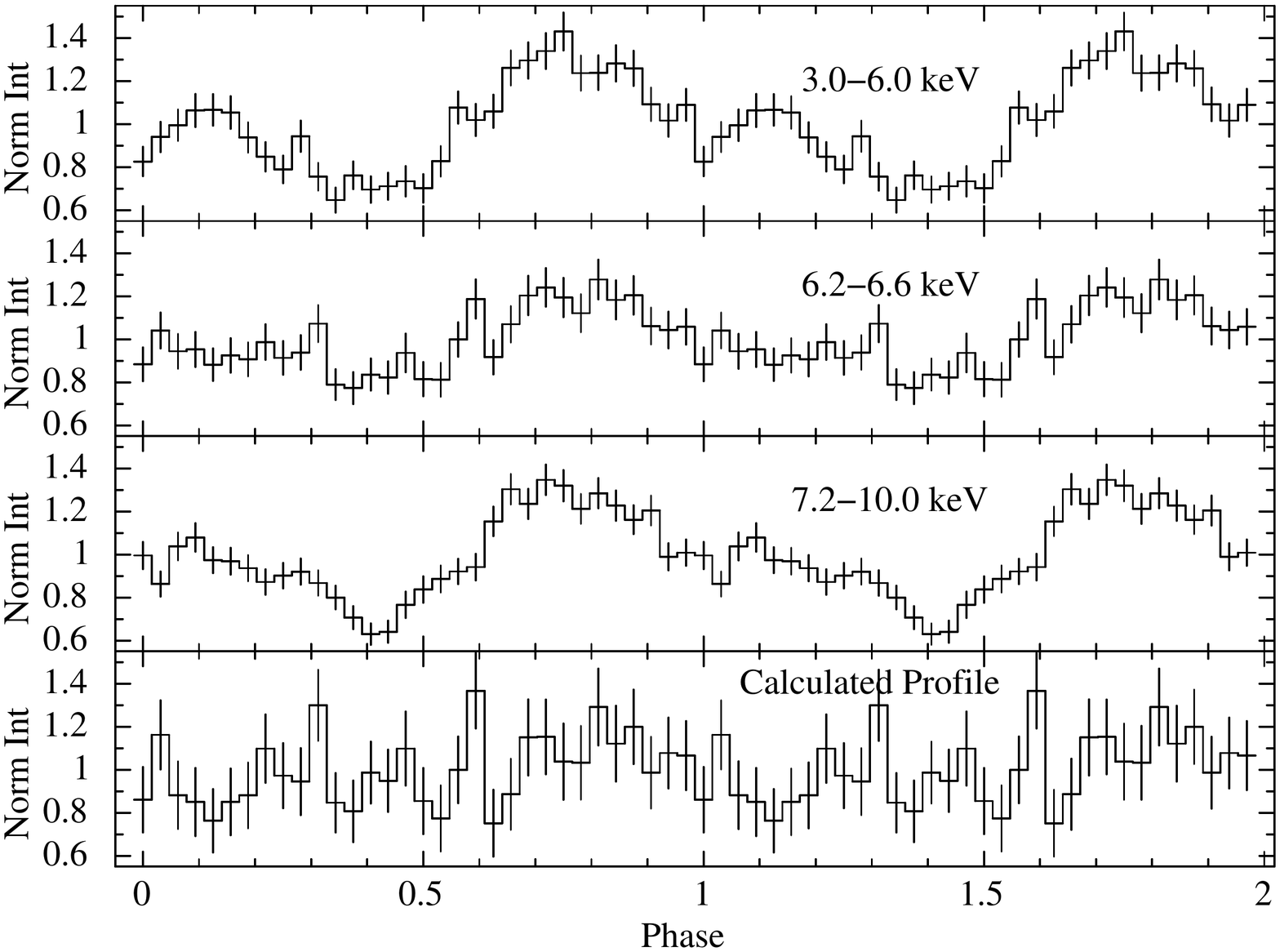}
    \caption{Energy resolved pulse profiles from observation \#6 (Left) and \#9 (Right). Top three panels show pulse profiles from low energy band (3.0-6.0 keV), mid energy band (6.2-6.6 keV), and high energy band (7.2-10.0 keV) respectively.}
    \label{enrespp}
\end{figure}   

The X-ray spectrum of this source showed strong Fe emission lines as discussed in the next section. To check the timing properties of this emission feature, we extracted pulse profiles in three different energy bands, viz. the low energy band (3.0-6.0 keV), the mid-energy band around Fe $K_{\alpha}$ line emission (6.2-6.6 keV), and the high energy band (7.2-10.0 keV). Such energy-resolved pulse profiles are shown in Figure \ref{enrespp} for observations \#6 and \#9. As seen in the figure, there is a significant amount of pulsation seen in the mid energy band. Mid energy band pulse profile has a contribution from the Fe $K_{\alpha}$ photons and the continuum photons in that energy band. We calculated the pulse profile of only the Fe $K_{\alpha}$ photons by removing the contribution of continuum photons. Since the continuum pulse profile in the band below and above the line energy are nearly identical, we assumed the pulse profile of continuum photons in the 6.2-6.6 keV energy band to be the average of the two other profiles. We subtracted the same from 6.2-6.6 keV energy band pulse profile to obtain the pulse profile of the Fe $K_{\alpha}$ photons. We assumed that the mid energy profile is the weighted average of Fe $K_{\alpha}$ photons pulse profile and continuum photons pulse profile, weights being equal to the proportion of flux they contribute to this energy band. The normalized pulse profile in the three energy bands, as shown in Figure \ref{enrespp} have been used for the same. \\

$I_{Avg}(n)$ is calculated from the normalized pulse of the low and high energy pulse profile as :
\begin{equation}
 \begin{aligned}
 I_{Avg}(n) = \frac{I_{Low}(n) + I_{High}(n)}{2}
\end{aligned}
\end{equation}

Here we have assumed that pulsation behavior of continuum in whole \xmm\ band is same. Intensity in $n^{th}$ bin due to Fe $K_{\alpha}$ line photons $I_{Line}(n)$ is given as:

\begin{equation}
\begin{aligned}
 I_{Line}(n) = 1 + \frac{BW+EQW}{EQW}\Bigg( \Big(I_{Mid}(n) -1 \Big) - \Big(1 - \frac{EQW}{BW+EQW} \Big)\Big( I_{Avg}(n) -1 \Big) \Bigg) 
\end{aligned}
\end{equation}

In this equation EQW is Equivalent Width of Fe $K_{\alpha}$ line emission, BW is total band width of mid energy range. These Fe $K_{\alpha}$ photons pulse profiles are plotted in bottom panels in Figure \ref{enrespp}. We do not observe any pulsed behavior by Fe $K_{\alpha}$ photons.\\

\subsection{Spectroscopy}

\begin{figure}
    \centering
    \includegraphics[width=0.48\textwidth]{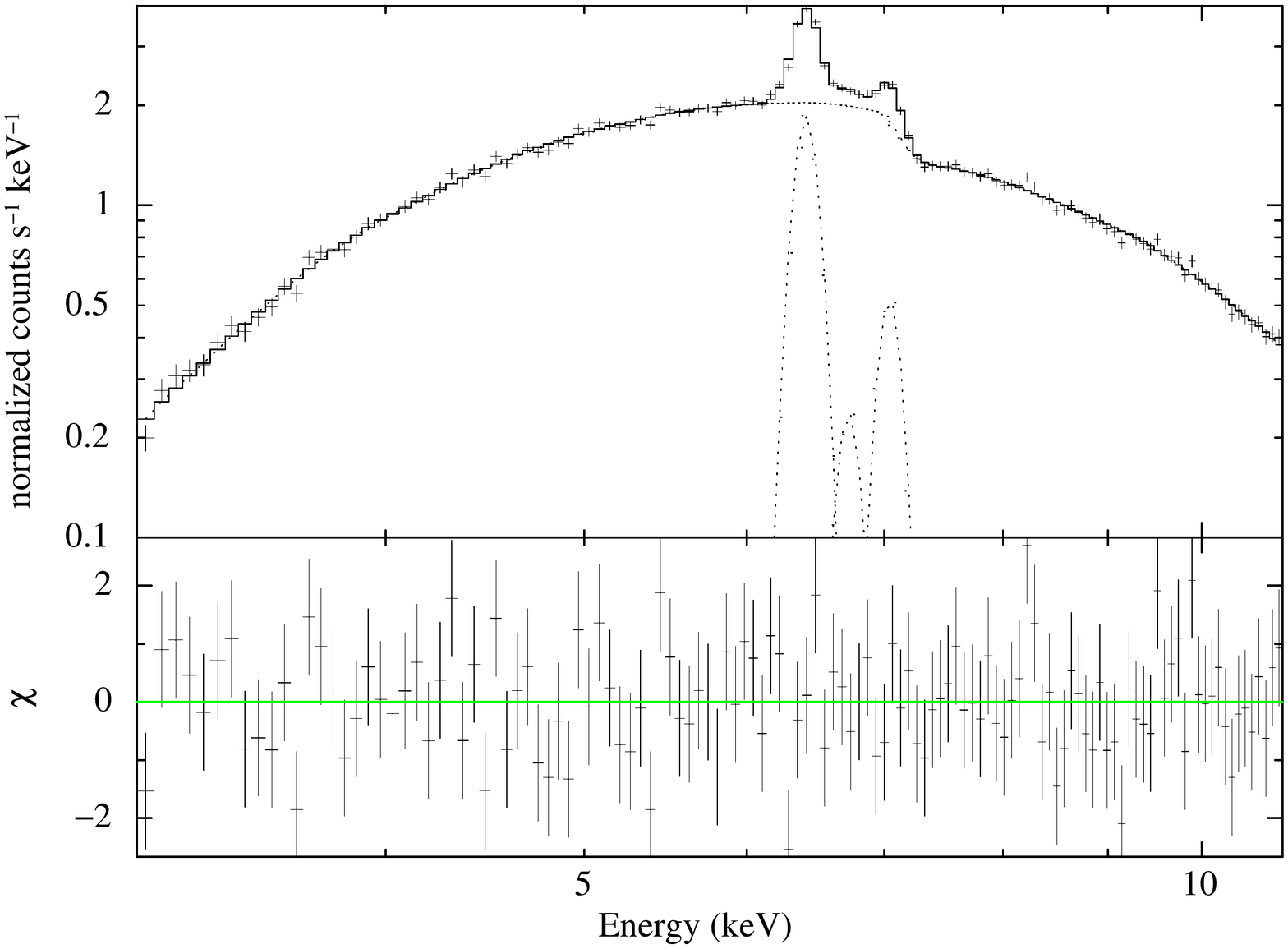}
    \includegraphics[width=0.48\textwidth]{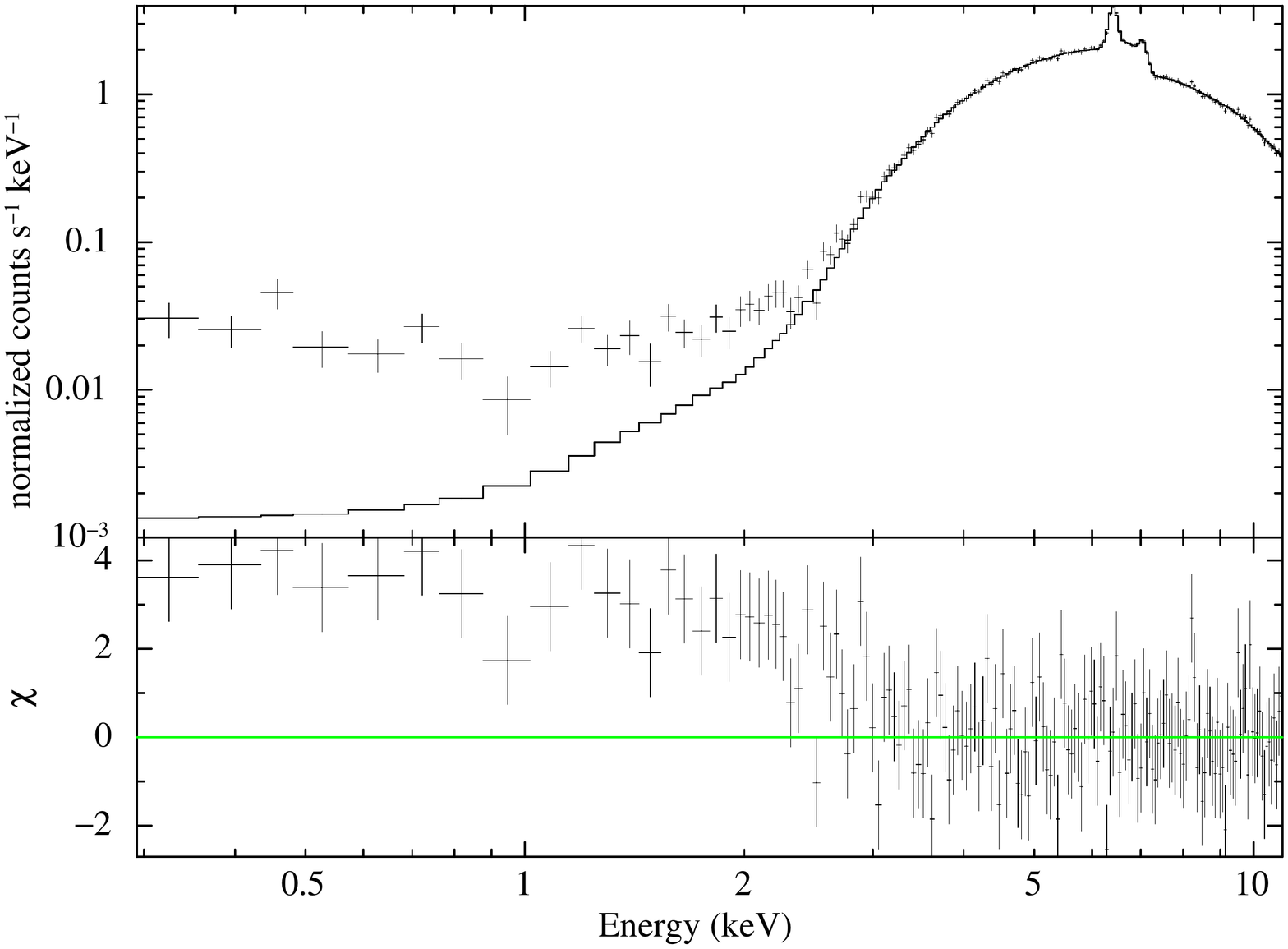}
    \caption{Spectrum of IGR J16320-4751 with a model fit from observation 8. \emph{Left panel:} Spectrum in range 3.0-11.0 keV. \emph{Right panel:} Spectrum in range 0.3-11.0 keV. Bottom panels shows residuals of fitted model.}
    \label{spec}
\end{figure}

The continuum spectrum of \src\ can be modeled either with a phenomenological powerlaw model \citep{2018A&A...610A..50P} or with a thermally comptonized  model (COMPTT, \citet{1994ApJ...434..570T}) to interpret the data in terms of physical conditions close to the surface of NS \citep{2018A&A...618A..61G}. In addition, a powerlaw with high energy cutoff model has been found to fit the spectral data as well \citep{2006MNRAS.366..274R,2018A&A...618A..61G}. We choose this model to describe the continuum in 3.0-11.0 keV energy range because it gives a better value of test statistic $\chi^2$ compared to the simple powerlaw model. We used 3 Gaussian components for line-like features in 6.0-7.5 keV energy range in a similar approach followed by \citet{2018A&A...618A..61G} by freezing their width to zero. The use of 3 Gaussian components was chosen compared to 2 Gaussian components as it improves the residuals. Only one Gaussian component was needed in observation 1 fit and two Gaussian components were required for 4 observations (see Table \ref{lpt}). While this describes the spectral data well at energies $>$3 keV for all the observations (shown in the left panel of Figure \ref{spec}), at lower energies a soft excess is seen (right panel of Figure \ref{spec}) in the some of the \xmm\  observations. To model this soft excess we tried three components viz: 1. additional absorption with partial covering 2. unabsorbed bremsstrahlung and 3. absorbed black-body. Neither of these models could account for the soft excess completely. A partial covering absorption component either gives unacceptable $\chi^2$ for some observations or large residuals for other observations. The bremsstrahlung model results in unreasonably high temperature ($\sim$200 keV) for the electron plasma. The blackbody model parameters are not well constrained when describing the soft excess data.  To quantify the flux of the soft excess component, we used the unabsorbed blackbody model and notice the value of the norm of this spectral component and convert it into flux. We detect the presence of soft excess in 6 out of 11 observations (see Table \ref{cpt}). The flux of this emission is in the range of (0.4-2.4)$\times10^{-13}$ erg cm$^{-2}$ s$^{-1}$. For other observations, this soft excess flux is consistent with zero. \\ %Table~\ref{lpt} list the soft excess flux values from different observations. \\  

\begin{table}
\centering
  \caption{Table of parameter values for the continuum component of spectral model.}
  \label{cpt}
  \begin{tabular}{||c|c|c|c|c|c|c|c||}
  \hline
   Obs. No. & \multicolumn{7}{|c|}{Parameters}  \\
  \hline
     &  $N_{H}$             & $E_{cut}$(keV)         & $E_{fold}$(keV)     & $\Gamma $              & Norm\tnote{2}          & Soft Excess           & $\chi^{2}_{red}$ \\
  \hline
  1  & $19^{+3}_{-6}$       & $8^{+1}_{-2}$          & $7^{+8}_{-1}$       & $0.8^{+0.3}_{-0.9}$    & $3^{+2}_{-2}$          &          -            &1.19 \\
  2  & $15.0^{+0.3}_{-0.9}$ & $7.4^{+0.2}_{-0.3}$    & $15^{+2}_{-1}$      & $0.43^{+0.05}_{-0.14}$ & $3.4^{+0.3}_{-0.7}$    & $0.9^{+1.1}_{-0.4}$   &1.21 \\
  3  & $27^{+1}_{-3}$       & $8.1^{+0.4}_{-0.6}$    & $11^{+4}_{-3}$      & $0.5^{+0.1}_{-0.1}$    & $4.6^{+0.1}_{-0.9}$ & $0.4^{+1.4}_{-0.2}$      &1.32 \\
  4  & $19.9^{+0.5}_{-1.0}$ & $8^{+1}_{-1}$          & $29^{+16}_{-9}$     & $0.39^{+0.09}_{-0.38}$ & $7^{+1}_{-1}$          & $0.6^{+1.3}_{-0.2}$   &1.13\\
  5  & $19^{+1}_{-1}$       & $7.0^{+0.9}_{-0.6}$    & $18^{+6}_{-5}$      & $0.3^{+0.2}_{-0.2}$    & $5^{+2}_{-2}$          &          -            &1.17 \\
  6  & $22^{+3}_{-2}$       & $7.8^{+0.4}_{-0.5}$    & $14^{+5}_{-3}$      & $0.43^{+0.09}_{-0.11}$ & $4.6^{+0.4}_{-0.9}$    &          -            &0.94 \\
  7  & $21^{+1}_{-1}$       & $7.6^{+0.6}_{-0.7}$    & $9^{+4}_{-3}$       & 0.43                   & $3.8^{+0.2}_{-0.2}$    &          -            &0.65 \\
  8  & $20.4^{+0.5}_{-0.3}$ & $7.7^{+0.2}_{-0.6}$    & $16^{+2}_{-2}$      & $0.24^{+0.06}_{-0.12}$ & $6.3^{+0.8}_{-1.3}$    & $2.4^{+3.4}_{-0.1}$   &0.92 \\
  9  & $46^{+1}_{-1}$       & $8.7^{+0.4}_{-0.5}$    & $14^{+6}_{-5}$      & 0.24                   & $3.6^{+0.1}_{-0.1}$    &          -            &1.16 \\
  10 & $20^{+1}_{-2}$       & $7.2^{+0.9}_{-0.6}$    & $10^{+3}_{-2}$      & $0.3^{+0.2}_{-0.2}$    & $4^{+2}_{-1}$          & $1.0^{+3.6}_{-0.6}$   &0.91 \\
  11 & $18^{+1}_{-4}$       & $7.2^{+0.3}_{-0.7}$    & $16^{+2}_{-1}$      & $0.3^{+0.1}_{-0.5}$    & $4.0^{+2.0}_{-0.9}$    & $0.4^{+0.3}_{-0.1}$   &1.13 \\
  \hline
  \end{tabular}
  \begin{tablenotes}
  \item {\bf Notes}: Equivalent hydrogen column $N_{H}$ is in units of $10^{22}$ atoms cm$^{-2}$. The normalization of photon power law is in units of photons $10^{-3}$ keV$^{-1}$ cm$^{-2}$ s$^{-1}$ at 1 keV. The soft excess in  0.3-3.0 keV energy range is given in the units of $10^{-13}$ erg cm$^{-2}$ s$^{-1}$. 
  \end{tablenotes}
\end{table}
 
\begin{table}
    \centering
    \caption{Table of parameter values for the line components of spectral model.}
    \label{lpt}
    \begin{tabular}{||c|c|c|c|c|c|c||}
    \hline
    Obs. No. & \multicolumn{6}{|c|}{Parameters} \\
    \hline
        & $E_{L1}$ (keV)           & $Norm_{L1}$          & $E_{L2}$ (keV)         &  $Norm_{L2}$         & $E_{L3}$ (keV)             &  $Norm_{L3}$        \\
    \hline 
    1   & $6.38^{+0.09}_{-0.11}$   & $ 0.3^{+0.2}_{-0.2}$  &         -              &         -            &          -                 &      -              \\
    2   & $6.398^{+0.002}_{-0.007}$& $1.78^{+0.09}_{-0.10}$& $6.68^{+0.07}_{-0.07}$ & $1.41^{+0.8}_{-0.8}$ &  $7.04^{+0.03}_{-0.03}$    & $2.7^{+0.8}_{-0.7}$ \\
    3   & $6.414^{+0.018}_{-0.004}$& $4.5^{+0.3}_{-0.3}$   &        -               &       -              &  $7.06^{+0.03}_{-0.04}$    & $7^{+2}_{-2}$       \\
    4   & $6.414^{+0.005}_{-0.011}$& $6.8^{+0.4}_{-0.4}$   & $6.75^{+0.10}_{-0.08}$ & $5^{+3}_{-3}$        &  $7.05^{+0.04}_{-0.04}$    & $12^{+3}_{-3}$      \\
    5   & $6.430^{+0.025}_{-0.006}$& $5.2^{+0.4}_{-0.4}$   & -                      & -                    &  $7.05^{+0.05}_{-0.06}$    & $7^{+3}_{-3}$       \\
    6   & $6.412^{+0.007}_{-0.023}$& $3.1^{+0.3}_{-0.6}$   & $6.62^{+0.08}_{-0.23}$ & $4^{+2}_{-2}$        &  $7.02^{+0.05}_{-0.04}$    & $6^{+2}_{-3}$       \\
    7   & $6.40^{+0.02}_{-0.02}$   & $2.4^{+0.5}_{-0.5}$   &  -                     & -                    &  $7.0^{+0.1}_{-0.2}$       & $5^{+4.0}_{-4}$     \\
    8   & $6.418^{+0.005}_{-0.016}$& $8.5^{+0.3}_{-0.3}$   & $6.73^{+0.02}_{-0.04}$ & $11^{+3}_{-3}$       &  $7.036^{+0.028}_{-0.007}$ & $26^{+3}_{-3}$      \\
    9   & $6.418^{+0.006}_{-0.002}$& $10.0^{+0.5}_{-0.5}$  & $6.73^{+0.04}_{-0.06}$ & $10^{+3}_{-3}$       &  $7.06^{+0.02}_{-0.03}$    & $19^{+4}_{-2}$      \\
    10  & $6.410^{+0.007}_{-0.004}$& $3.8^{+0.4}_{-0.3}$   & -                      & -                    &  $7.00^{+0.06}_{-0.05}$    & $7^{+3}_{-3}$       \\
    11  & $6.42^{+0.01}_{-0.04}$   & $3.9^{+0.4}_{-0.3}$   & $6.66^{+0.08}_{-0.06}$ & $4^{+4}_{-3} $       &  $7.03^{+0.02}_{-0.03}$    & $9^{+3}_{-3}$       \\
    \hline
    \end{tabular}
    \begin{tablenotes}
    \item {\bf Notes}: The normalisation of first Gaussian is given in the units of $10^{-4}$ photons cm$^{-2}$ s$^{-1}$. The normalisation of second and third Gaussian's is given in the units of $10^{-5}$ photons cm$^{-2}$ s$^{-1}$. 
    \end{tablenotes}
\end{table} 

\begin{figure}
    \centering
    \includegraphics[width=0.49\textwidth,trim=1cm 2cm 0cm 0cm,clip=true]{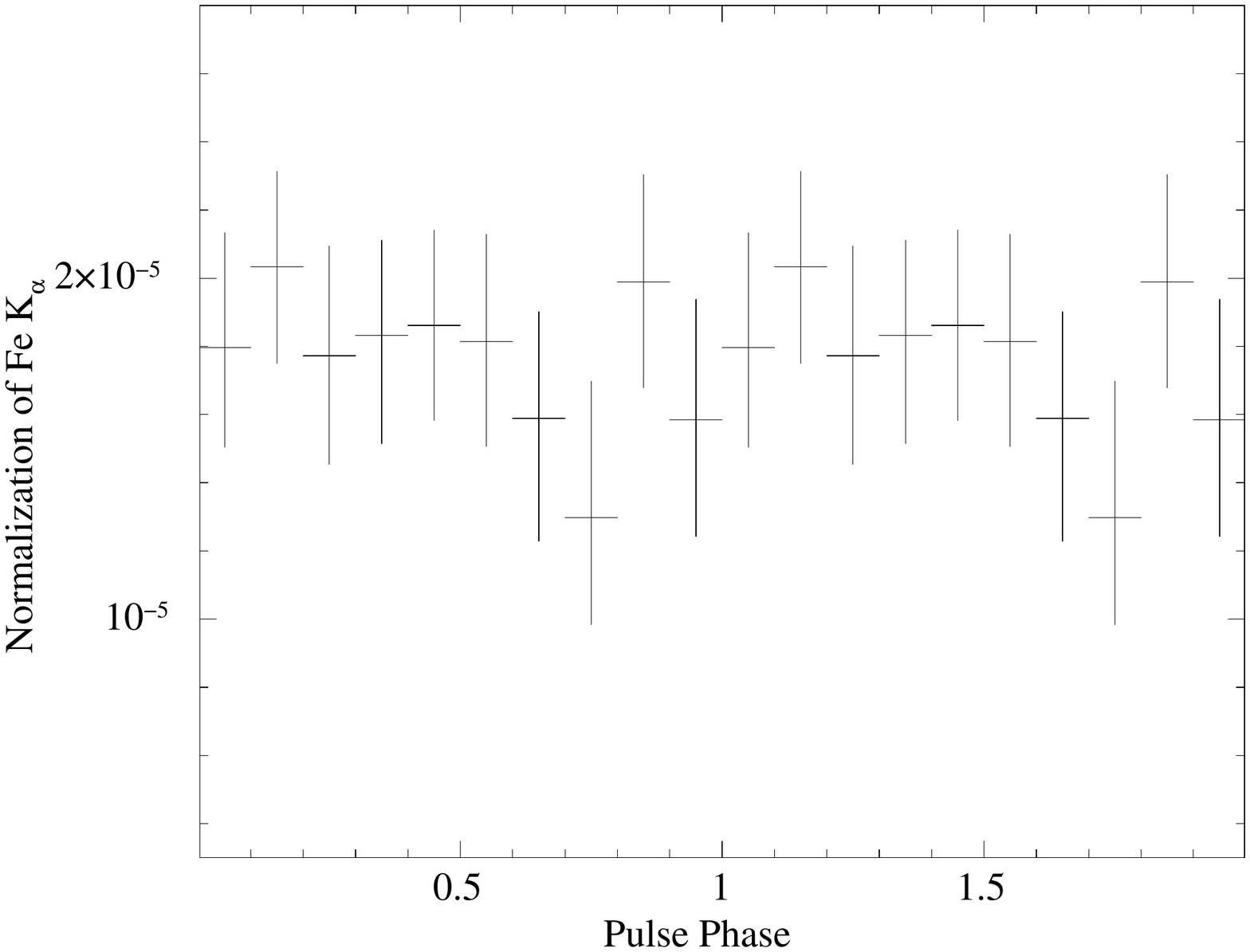}
    \includegraphics[width=0.44\textwidth,trim=1cm 0cm 0cm 0cm,clip=true]{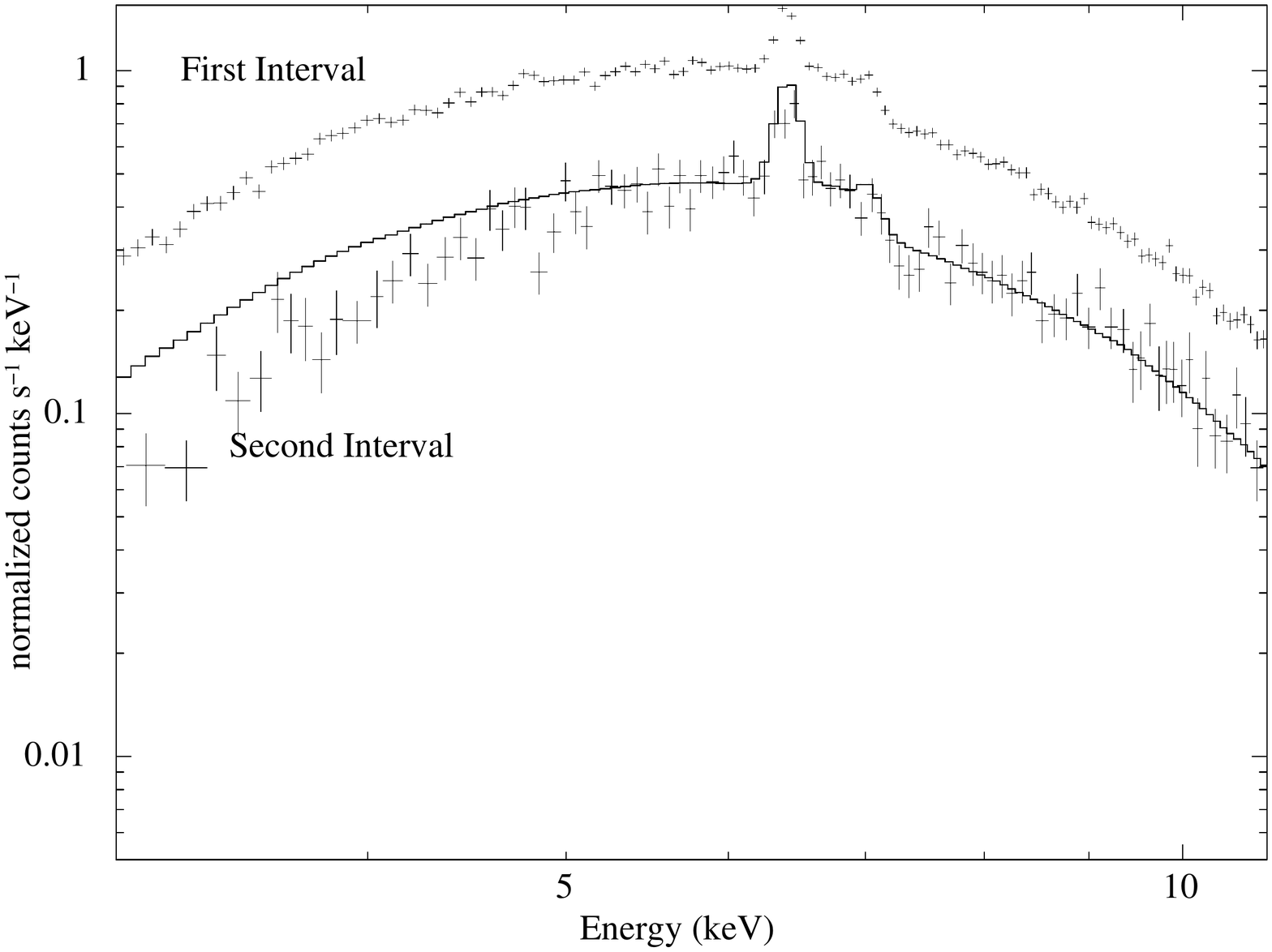}
    \caption{\emph{Left panel:}  The behavior of normalizaion of Fe $K_{\alpha}$ line in the spectra of \src\ with pulse phase. \emph{Right panel:} Spectra of \src\/ from two adjescent intervals in the light curve of observation \#2. Solid line represents the model of the first interval rescaled to the flux level of the second interval spectra.}
    \label{fig:cosp}
\end{figure}

The behavior of Fe $K_{\alpha}$ photon with pulse phase can also be probed using pulse phase-resolved spectroscopy. We checked for variation of the flux of the Fe $K_{\alpha}$ line in the phase-resolved spectra of two observations: one with the largest exposure time (observation \#2) and another showing the largest equivalent width of the Fe $K_{\alpha}$ line in phase averaged spectrum (observation \#8). We extracted ten phase-resolved spectra with a phase bin width of 0.1 from these observations. We fit all the phase-resolved spectra with the same model as used for the phase averaged spectrum. Line energies for three Gaussian's were frozen at phase averaged value and line widths were fixed to zero for all the phase averaged spectra. We noted the norm of the Gaussian corresponding to the Fe $K_{\alpha}$ line from all the phase resolved spectra. We did not find any statistically significant variation in the value of this spectral parameter with phase for both the observations (See the left panel of Figure \ref{fig:cosp}). Hence, this additional check also validates that iron line photons do not show any variation with the pulse phase. We discuss the implications of such behavior on the distribution of matter.\\ 

We performed a time-resolved spectral analysis to understand the physical reason behind the peculiar pulse to pulse variations in the light curves of \src. We selected an interval in observation \#2 where the pulsation from the source disappears for two consecutive cycles and another interval of the same duration just before this interval where the pulsation is seen clearly. We accumulated the phase-averaged source and background spectra in 3.0-11.0 keV from these intervals. In the right panel of Figure \ref{fig:cosp}, we show the spectra from these two intervals. We fitted both these spectra simultaneously with an absorbed powerlaw with high energy cutoff plus three Gaussians (same spectral model as in case of spectral data from the full observation). The parameters values of second and third Gaussians for second interval were tied to the values of these components in the first interval. Between the two intervals, there is a flux change by a factor of two or more. From the spectral fitting, we found an increase in column density of hydrogen from $13.3^{+0.7}_{-1.1}\times10^{-22}$ cm$^{-22}$ in the first interval to $22^{+3}_{-2}\times10^{-22}$ cm$^{-22}$ in the second interval. The increase in column density becomes visually apparent when we rescale the spectral model of the first interval to the flux level of the second interval. In addition, the equivalent width of 6.4 keV emission line increased from 108 eV in the first interval to 143 eV in the second interval. Although the spectral index of the powerlaw component was found match within error bars for both the intervals. We shall discuss the implication of these spectral changes in the discussion section.

\section{Discussion}\label{sec:dis}

We report the results from a detailed timing and spectral study of HMXB pulsar \src\ using all available archival \xmm\ observations taken between 2003 and 2008. We also use a long term \bat\ light curve between 2005 and 2020. Archival \xmm\ observations were used by \cite{2018A&A...618A..61G} to study the orbital variation of X-ray emission and column density. They reported high variability and flaring activity in this source. Also, they derived a correlated evolution of column density and flux of the Fe $K_{\alpha}$ along the orbit of this system. Our analysis brought some new results about source behavior. Specifically, we brought up three interesting observations. These observations are to do with 1) the timing variability in terms of pulse to pulse variations in different light curves and average pulse profile variability, 2) the pulse phase behavior of Fe $K_{\alpha}$ photons, and 3) the presence of soft excess below 3 keV in the spectra of the source. From the \bat\ data of this source, we found an orbital period of $8.991\pm0.001$ days which is more accurate than previously reported values \citep{2018A&A...618A..61G,2005ATel..649....1C}. We discuss the implications of these results in the following section below.

\subsection{Timing variability and variable pulse profile}

The light curves of \src\/ are highly variable. This variability is in terms of the difference in average count rates from different \xmm\/ observations, flaring activity, and sudden changes in count rate. This seemingly random variation is characteristic of absorption and accretion from clumpy wind given by companion \citep{2019ApJ...883..116P}. In addition to this, pulse to pulse variations are noticed in 6 out of 11 \xmm\/ observations. There are several instances of pulse to pulse variation for single and multiple pulses together. During such instances, either pulse is missing or the pulse shape drastically changed. Spectral comparison of one such instance where good spectral data was found with an adjoining interval with clear pulsation indicate that these changes are driven by variable absorption in \src\/. The increase in the absorption column density and equivalent width of the 6.4 keV emission line indicates that NS is passing through a dense clump of matter when the pulsation is missing in the light curve. Probably the scattering of soft X-ray photons by the matter clump causes the smearing of X-ray pulsation. Such instances are more common in BeXRB sources like \gxt\/, \aof\/, \gro\/, and \voz\/ due to highly variable winds from the companion but SgXRBs \vel\ and \oao\ has also been found to show such a behavior \citep{2014EPJWC..6406012K,2014MNRAS.442.2691P}. Our analysis of \src\/ indicates that such variations can also be observed in highly obscured SgHMXB sources.  The strong timing variability of \src\/ leads to variation in the average pulse profile from different observations even when taken a few days apart. Variability is in pulse shape as well as in pulse fraction. This makes it difficult to estimate the pulse arrival time delays using the correlation of pulse profiles. A mass function of $\sim$280 $M_{\odot}$ or orbital period derivative of 5$\times$10$^{-4}$ ss$^{-1}$  determined by using these observations seems unlikely and large phase arrival delays are driven by intrinsic variability of pulse profiles. If this is indeed true, hard X-ray pulse profiles and phase lags from NuSTAR/AstroSat observations can possibly give the orbital solution in this source.

\subsection{Pulse phase behaviour of the iron line}

Fe $K_{\alpha}$ emission lines are ubiquitous features in the spectra of X-ray binaries with a neutron star. This spectral feature is caused by reprocessing of hard X-rays from the NS from neutral or partially ionized material in the system. Candidates for the reprocessing site are the Alfven shell, the accretion disk (in those sources powered by the Roche lobe overflow), accretion wakes, the stellar wind, and the surface of the companion star. Our energy resolved pulse profile revealed a pulsating behavior in 6.2-6.6 keV energy band. However, the true behavior of Fe $K_{\alpha}$ photons (obtained by subtracting the proper continuum emission contribution from 6.2-6.6 keV energy band) is non-pulsating. This is also validated with the pulse phase resolved spectroscopic analysis. Such a non-pulsating behavior of Fe $K_{\alpha}$ emission has been reported in some other HMXBs like GX 1+4 \citep{1999ApJ...510..369K}, Vela X-1 \citep{2002aprm.conf..355P}, and 4U 1901+03 \citep{2009ApJ...707.1016L}. This result rules out material having an anisotropic distribution as in the case of Alfven shell, accretion stream, the surface of the companion star, or a warped structure in the accretion disk. Line energy of the emission line in 6.2-6.6 keV band obtained from spectral fitting of different observations lies between 6.38-6.43 keV (see Table \ref{lpt}). These values correspond to either a low or a moderate ionization state for Fe atoms \citep{1969ApJS...18...21H}. The highest value of line energy corresponds to Fe XVII. Hence, the line emitting material is also not located very close to NS. The distance between the line emitting region and NS can be estimated using the formula for the ionization parameter,  $\zeta$ = L/($N_{H}\times r$), where L is the luminosity of the source, $N_H$ = $\rho \times r$ is column density in the line of sight towards the source. From the line energy of the emission line in 6.2-6.6 keV band, we can estimate $log \zeta < 2.1$ (Figure 5. of \cite{2004ApJS..155..675K}), therefore $\zeta < 125$. The unabsorbed bolometric flux in 0.01-100 keV has been estimated to be $4.4\times10^{-10}$ erg $cm^{-2}$ $s^{-1}$ using a combined spectral fit of the \xmm\ and \integral\ \citep{2006MNRAS.366..274R}. This leads to a luminosity of $6.4\times10^{35}$ erg $s^{-1}$ for a distance of 3.5 kpc estimated for this source. Using these values of L and $\zeta$, we get $r>$1.5-4.3$\times10^{10}$ cm (corresponding to $N_{H}$=1.2-3.5$\times10^{23}$ $cm^{-2}$ obtained from spectral fits) for the distance between NS and line emitting region. An upper bound on the distance of the line emitting region comes from the fact that even a symmetrically distributed material at a large distance can produce pulsed emission due to difference in light travel time. For $P_{spin} = 1305.9$ sec, we get $r<4\times10^{13}$ cm for line emission to remain unpulsed. The Fe $K_{\alpha}$ line emitting material is located between these two distance bounds in this source. 

\subsection{Presence of soft excess}

\begin{figure}
    \centering
    \includegraphics[scale=0.28]{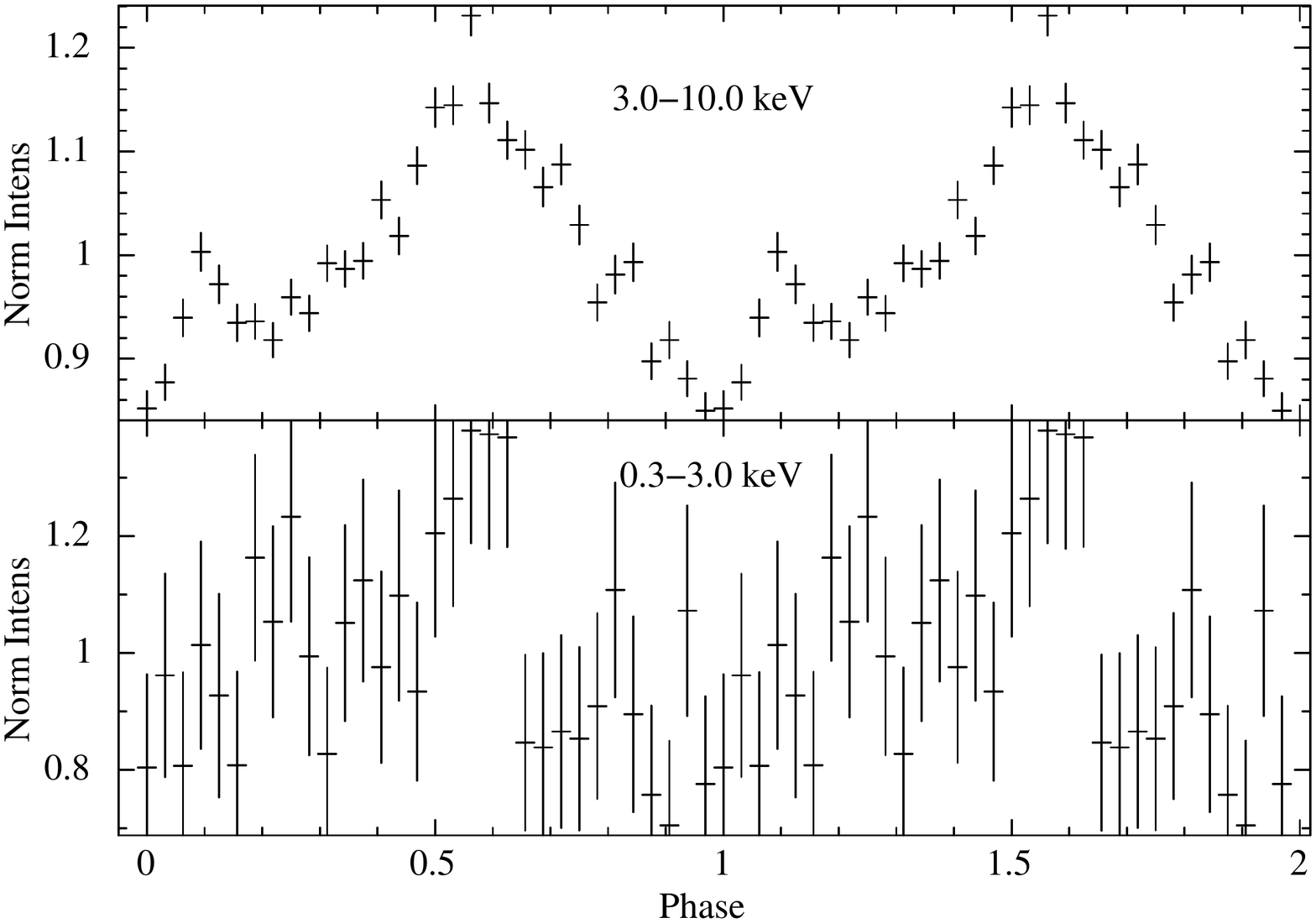}
    \includegraphics[scale=0.28]{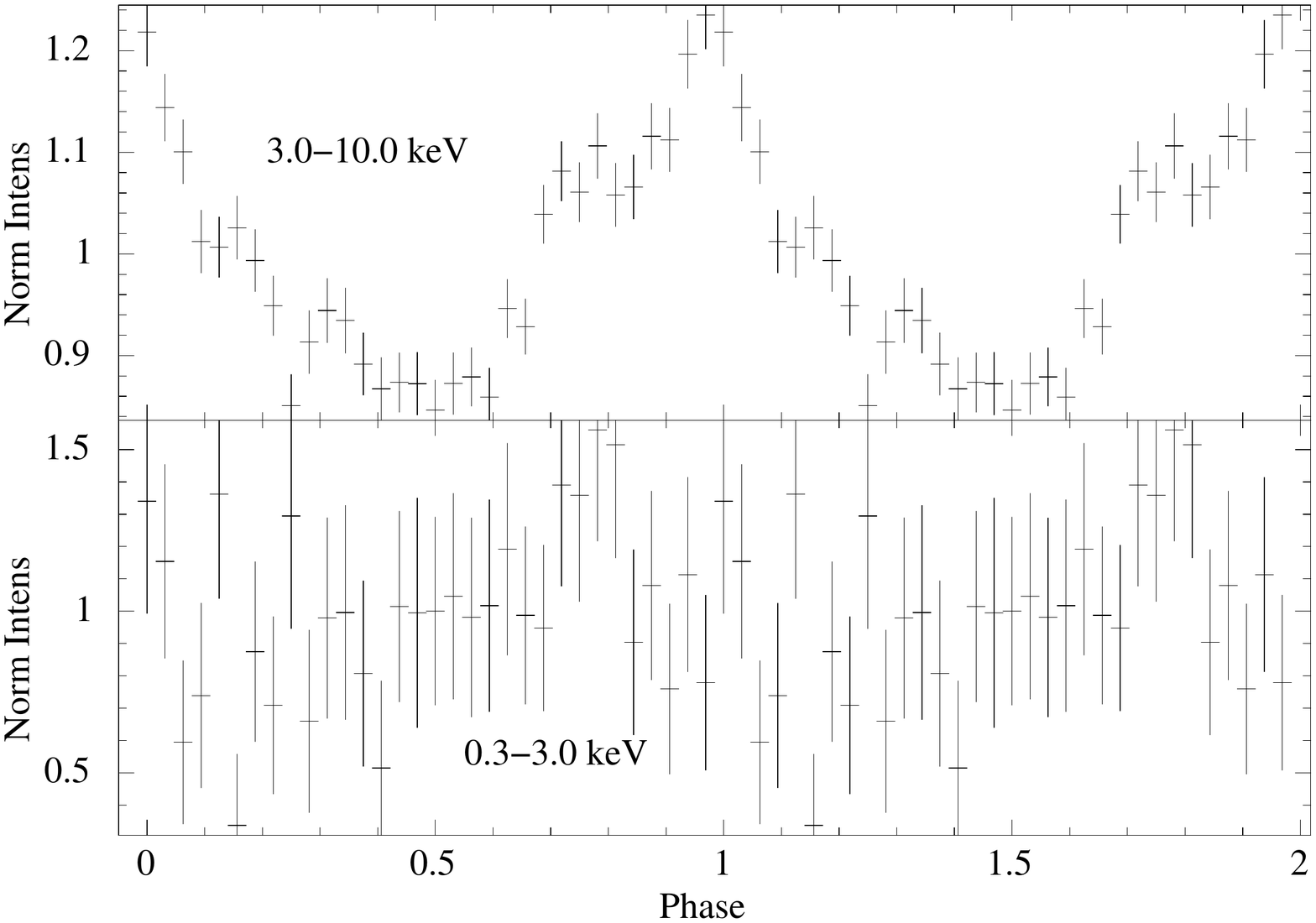}
    \caption{Comparison of pulse profile in soft excess band (0.3-3.0 keV) with the higher energy band (3.0-10.0 keV) for observation \#8 and \#10 which show strong soft excess emission.}
    \label{fig:sep}
\end{figure}

Soft excess emission is a spectral feature of excess intensities over the extrapolation of a power law fitted to higher energies. Such emission is seen in many HMXBs and is explained by 4 possible mechanisms \citep{2004ApJ...614..881H}: (1) emission from the accretion column (2) reprocessing by optically thick, dense material around the NS (3) thermal emission by collisionally energized diffuse gas around the NS (4) reprocessing of hard X-rays by diffuse material around NS. The soft excess emission in the non-flaring duration of observation \#2 was fitted using a blackbody component with a temperature of 0.07 keV \citep{2006MNRAS.366..274R}. They suggested that the soft excess emission is due to collisionally energized cloud based on the energy conservation argument \citep{2006MNRAS.366..274R}. However, the blackbody component does not give a good fit for the other observations where such emission is observed. Another model component of thermal Bremsstrahlung gives a good fit but the value of plasma temperature seems unusually high ($\sim$200 keV) for such an environment. Other sources like \smx\ and \lmx\ show plasma temperatures of $\sim$0.1 keV \citep{2002ApJ...579..411P}. A common feature among all observations of this source showing soft excess is the absence of pulsation in the 0.3-3.0 keV band (see Figure \ref{fig:sep}). The absence of pulsation combined with the fact this source has very low luminosity suggests that the soft excess is emitted by diffuse gas in this source similar to other sources like Vela X-1, RX J0101.3-7211, and AX J0101-722 \citep{2004ApJ...614..881H}. Further, this could be either a collisionally ionized or photoionized gas. Reprocessing from a diffuse gas cloud has been found to contribute to soft excess either during the eclipse or just before the eclipse in sources like \vel\ \citep{2002ApJ...564L..21S}, 4U 1700-37 \citep{2003ApJ...592..516B}, and \fuf\ \citep{2019ApJS..243...29A}. High-resolution spectral studies in these sources have shown a large variety of emission features like fluorescent emission lines from the mid-Z elements and iron, radiative recombination continuum feature, and lines from H-like, He-like ions. Hence, the soft X-ray emission of \src\ should be studied with high-resolution spectral observation with \rgs\ or \hetg\ to confirm the nature of soft excess and complex composition of the emitting plasma. \\

\textit{Acknowledgements.} This work is based on observation data and software tools from \xmm\ (An ESA Science mission with instruments and contributions directly funded by ESA member states and USA) and NASA's High Energy Astrophysics Science Archive (HEASARC), a service of Goddard space flight center and the Smithsonian Astrophysical Observatory.

\bibliographystyle{apalike}
\bibliography{bibtex.bib}
\end{document}